\documentclass[12pt,noshowpacs,nofootinbib,notitlepage,amsmath,superscriptaddress]{revtex4-1}
\usepackage{setspace}
\usepackage{graphicx,psfrag}  
\usepackage{amsfonts}
\usepackage{graphicx,color}
\usepackage{appendix}
\usepackage{mathtools}

\newcommand{\x}{\mathsf{x}}

\DeclareMathOperator{\expo}{e}

\newcommand{\be}{\begin{equation}}
\newcommand{\ee}{\end{equation}}
\newcommand{\ba}{\begin{eqnarray}}
\newcommand{\ea}{\end{eqnarray}}
\newcommand{\bea}{\begin{equation}\begin{aligned}}
\newcommand{\eea}{\end{aligned}\end{equation}}
\newcommand{\baa}{\begin{array}}
\newcommand{\eaa}{\end{array}}

\DeclareMathOperator{\Phin}{\Phi^{\text{in}}_{\omega\ell}}
\DeclareMathOperator{\Phup}{\Phi^{\text{up}}_{\omega\ell}}

\DeclareMathOperator{\rin}{\mathsf{\rho^{\text{in}}_{\omega\ell}}}
\DeclareMathOperator{\rup}{\rho^{\text{up}}_{\omega\ell}}
\DeclareMathOperator{\psiL}{\mathsf{\psi^{\text{left}}_{\omega\ell}}}
\DeclareMathOperator{\psiR}{\psi^{\text{right}}_{\omega\ell}}

\newcommand{\Rin}{R^{\text{in}}_{\omega\ell}}
\newcommand{\Rup}{R^{\text{up}}_{\omega\ell}}
\newcommand{\Tup}{T^{\text{up}}_{\omega\ell}}
\newcommand{\Tin}{T^{\text{in}}_{\omega\ell}}
 
\begin{document}
 
\title{Unruh-DeWitt Detector Differentiation of Black Holes and Exotic Compact Objects \vspace{0.5cm}}
\author{Bob Holdom}
\email{bob.holdom@utoronto.ca}
\affiliation{Department of Physics, University of
Toronto, 60 St. George St., Toronto, Ontario, Canada M5S 1A7}

\author{Robert B. Mann}
\email{rbmann@uwaterloo.ca}
\affiliation{Department of Physics and Astronomy, University of Waterloo, Waterloo, Ontario, Canada, N2L 3G1}
\author{Chen Zhang}
\email{czhang@physics.utoronto.ca}
\vskip 0.5cm
\affiliation{Department of Physics, University of
Toronto, 60 St. George St., Toronto, Ontario, Canada M5S 1A7}
\affiliation{Department of Physics and Astronomy, University of Waterloo, Waterloo, Ontario, Canada, N2L 3G1}

\date{\today}

\begin{abstract}
We study the response of a static Unruh-DeWitt detector  outside an exotic compact object (ECO) with a general reflective boundary condition in 3+1 dimensions. The horizonless ECO, whose boundary is extremely close to the would-be event horizon, acts as a black hole mimicker.  We find that the response rate is notably distinct from the black hole case, even when the ECO boundary is perfectly absorbing. For a (partially) reflective ECO boundary, we find resonance structures in the response rate that depend on the different locations of the ECO boundary and those of the detector. We provide a detailed analysis in connection with the ECO's vacuum mode structure and  transfer function.
\end{abstract}

\maketitle

\section{Introduction}
The discovery of gravitational waves from compact binary mergers have inspired many recent studies on exotic compact objects (ECOs), 
whose defining feature is their compactness: their radius is very close to that of a  black hole with the same mass whilst lacking an event horizon.
Explicit examples of such compact horizonless objects include gravastars~\cite{Mazur:2001fv}, boson stars~\cite{Schunck:2003kk}, wormholes, and 2-2 holes~\cite{Holdom:2016nek}.  While some investigations are concerned with the consistency of semi-classical physics
near horizons \cite{Baccetti:2018otf,Terno:2020tsq},
most studies are concerned with finding distinctive signatures from postmerger gravitational wave echoes~\cite{Abedi:2016hgu, Mark:2017dnq,Conklin:2017lwb, Cardoso:2019rvt,Holdom:2019bdv}, in which a wave that falls inside the gravitational potential barrier  reflects off the ECO boundary and returns to the barrier after some time delay $t_d$. 

In this paper, we study the response of an Unruh-DeWitt (UDW) detector~\cite{unruh,deWitt} in this context. Assuming no angular momentum exchange, the
light-matter interaction is well described as a local interaction between  a scalar quantum field $\phi(x)$ and 
a local detector comprising two states: $\left|0\right\rangle$, with vanishing state energy, and $\left| \Omega \right\rangle$ of energy $\Omega$, where $\Omega$ may be a positive or negative real number \cite{Funai:2018wqq}. If $\Omega>0$ then $\left|0\right\rangle$ is the ground state. Known as a UDW detector, 
the interaction Hamiltonian is
\be
H_{\text{int}}=\lambda\chi(\tau)\mu(\tau)\phi(\x(\tau))\,,
\label{eq:techIntro:Hint} 
\ee
where $\chi(\tau)$ is the switching function encoding the time dependence of the interaction, whose coupling constant is   $\lambda$, and
 $\x(\tau)$ is the detector's trajectory with $\tau$  its proper time.  The quantity $\mu({\tau})$ is the monopole operator of the detector $\mu=e^{i\Omega\tau}\sigma^{+}+e^{-i\Omega\tau}\sigma^{-}$ with $\sigma^{+}=\left|\Omega\right\rangle \left\langle 0\right |$ and $\sigma^{-}=\left|0\right\rangle \left\langle \Omega\right |$   the respective raising and lowering operators.
 
Perturbing to first-order in $\lambda$, the transition probability is~\cite{byd}
\begin{equation}
P(\Omega)=c^2{|\langle0|\mu(0)|\Omega\rangle|}^2\mathcal{F}\left(\Omega\right)
\ , 
\label{eq:techIntro:prob}
\end{equation}
where $\mathcal{F}(\Omega)$ is the response function
\begin{equation}
\mathcal{F}(\Omega)=
 \int_{-\infty}^{\infty}\mathrm{d}\tau^{\prime} \,
\int_{-\infty}^{\infty}\mathrm{d}\tau^{\prime\prime} \,\chi(\tau^{\prime})\chi(\tau^{\prime\prime})\, 
\mathrm{e}^{-i \Omega (\tau^{\prime}-\tau^{\prime\prime})}\, W(\tau^{\prime},\tau^{\prime\prime})\,.
\label{eq:techIntro:respfunc}
\end{equation}
 $W(\tau',\tau'')=\langle 0| \phi(\mathsf{x}(\tau')) \phi(\mathsf{x}(\tau''))|0\rangle $ is the pull-back of
the Wightman distribution on the detector's
worldline. When the switching function is a Heaviside step function whose
 switch-on time is taken to be in the asymptotic past, we can  drop the external $\tau^{\prime}$-integral
of~\eqref{eq:techIntro:respfunc} by a change of variables and obtain the
transition rate~\cite{byd,Louko:2007mu}
\begin{equation}
\mathcal{\dot{F}}\left(\Omega\right)= \int^{\infty}_{-\infty}\,\mathrm{d}s\,\mathrm{e}^{-i\Omega s}\, W(s)\,,
\label{eq:transRate}
\end{equation}
representing the number of particles detected per unit proper time.
 
The spacetime of  a  Schwarzschild black hole in $3+1$ spacetime dimensions has the metric form (in geometric units where $c=G=1$)%
\begin{equation}
ds^2=-\left(1-\frac{2M}{r}\right)dt^2+\left(1-\frac{2M}{r}\right)^{-1}dr^2+r^2\left(d\theta^2+\sin^2{\theta}d\phi^2\right),
\end{equation}
where $M$ is the black hole mass, with the event horizon at $r_H=2M$.
Mode solutions of the Klein-Gordon equation for the massless scalar field $\phi$ in this background have the form
\begin{equation}
\phi=\frac{1}{\sqrt{4\pi\omega}} \frac{\rho_{\omega\ell}(r)}{r} Y_{\ell m}\left(\theta,\phi\right)\expo^{-i\omega t},
\end{equation}
where $Y_{\ell m}$ is the spherical harmonic function. The radial function $\rho_{\omega\ell}$ satisfies 
\begin{equation}
\frac{d^2 \rho_{\omega\ell}}{dr^{*2}}+\left\{\omega^2-V(r)\right\}\rho_{\omega\ell}=0 \,,
\label{eom:rho}
\end{equation}
where the tortoise coordinate $r^*$ satisfies $dr^*/dr=(1-2M/r)^{-1}$. The gravitational potential is
\be
V(r)=\left(1-\frac{2M}{r}\right)\left[\frac{\ell(\ell+1)}{r^2}+\frac{2M}{r^3}\right]
\label{Vr}
\ee
for the scalar field perturbation of angular momentum $l$.
$V(r)$ has a peak at  $r_{\rm peak}$, which is close to $r=3M$.  Note that we can rescale $\bar{r}=r/(2M),\,\bar{\omega}=2M\omega$ so that the dependence on $M$ in all equations is removed. 

We model the ECO as an exterior Schwarzschild spacetime patched to a spherically symmetric interior metric at radius $r =r_0$, with $r_0/r_H-1\ll 1$, with a boundary condition at $r_0$ parameterized by the reflection coefficient $R^W$. Typical quantum gravity arguments suggest that $r_0-r_H$ is related to the Planck length, as for the brick wall model~\cite{tHooft:1984kcu}. The space between  $r_0$ and $r_{\rm peak}$ acts as a cavity, and has an associated time delay\footnote{LIGO data favors the time delay $t_d\approx 600\,M$~\cite{Holdom:2019bdv}, which sets a preferred value for $r^*_0\approx-149.2$.}
\be
t_d \equiv 2 (r^*_{\rm peak}-r^*_0)
\ee
expressed in tortoise coordinates. An ECO boundary closer to the would-be horizon (i.e., smaller $r^*_0$) maps to a larger time delay $t_d$.  The tortoise coordinate can be chosen such that $r^{*}_{\rm peak}\approx0$.
\section{Formalism}

In this section, we review the formalism for the calculation of the detector's response rate in the black hole  context~\cite{Hodgkinson:2014iua}, which we will then extend to the ECO scenario.

\subsection{Black hole}

For a black hole, the gravitational potential vanishes at $r^*\to \pm \infty$ such that Eq.~(\ref{eom:rho}) has two simple asymptotic solutions: $\rho_{\omega l}^\textrm{up}(r^*\to \infty)=e^{ i\omega r^*}$,  $\rho_{\omega l}^\textrm{in}(r^*\to -\infty)=e^{-i\omega r^*}$. 
The normalized field modes
\begin{equation}
\begin{aligned}
\Phin &=T_{\omega l}^\textrm{in}\frac{\rho_{\omega l}^\textrm{in}}{r},\,\quad \Phup &=T_{\omega l}^\textrm{up}\frac{\rho_{\omega l}^\textrm{up}}{r}\, ,
\label{norm}
\end{aligned}
\end{equation}
satisfy the respective boundary conditions
\begin{eqnarray}
\Phi_{\omega l}^\textrm{in} r&\to&\left\{\begin{array}{cc}
\,T_{\omega l}^\textrm{in} e^{-i\omega r^*}, & r^*\to -\infty \\ 
e^{-i\omega r^*}+R_{\omega l}^\textrm{in}e^{i\omega r^*}, & r^*\to\infty\end{array}\right.\label{Phi_in_Asymtotic} \\
\Phi_{\omega l}^\textrm{up} r&\to&
\left\{\begin{array}{ll}
R_{\omega l}^\textrm{up} \,e^{-i \omega r^*}+e^{i \omega r^*},& r^*\to -\infty\\
T_{\omega l}^\textrm{up} \,e^{i \omega r^*},& r^*\to\infty\end{array} 
\right.    \label{Phi_up_Asymtotic}
\end{eqnarray}
such that $\Phi_{\omega l}^\textrm{in}$ represents a wave mode incident on the potential barrier from spatial infinity, giving rise to a reflected wave of amplitude $R_{\omega l}^{\rm in}$ and to a transmitted wave of amplitude $T_{\omega l}^\textrm{in}$. Conversely, $\Phi_{\omega l}^\textrm{up}$ represents a wave incident from the past event horizon and is partially reflected by a factor of $R_{\omega l}^{\rm up}$ and partially transmitted with an amplitude $T_{\omega l}^\textrm{up}$. 

From the Wronskian relations, one can derive 
\bea
T_{\omega l}^\textrm{in}&=\frac{2i\omega}{ W[\rin,\rup]} , \,\,\, \Rin=-\frac{W[\rin,\rho^{\text{up}\,*}_{\omega \ell}]}{W[\rin,\rup]},
\label{Tin_norm}
\eea
and\footnote{Note that some literature~\cite{Mark:2017dnq,Conklin:2017lwb,Cardoso:2019rvt} denote $\Tup$ as $T_{\rm BH}$,  and $\Rup$ as $R_{\rm BH}$.} 
\be
 \Tup=T_{\omega l}^\textrm{in}, \quad  ,\,\, \Rup=-\frac{\Tin}{T^{\text{in}*}_{\omega\ell} }  R^{\text{in}*}_{\omega\ell} 
\label{Tup_norm}
\ee
where $W$ denotes the Wronskian $W[a,b]=a b'- b a'$. 
Energy conservation furthermore implies  $|\Tup|^2+|\Rup|^2=1$, and $|\Tin|^2+|\Rin|^2=1$ . 

The Boulware state has positive-frequency modes with respect to the physical Schwarzschild  time $t$.  It reduces to the Minkowski vacuum at spatial infinity but induces a diverging stress-energy near the black hole horizon.
The quantum scalar field in this state can be expanded as
\begin{equation}
\begin{aligned}
\psi=\sum^{\infty}_{\ell=0}\sum^{+\ell}_{m=-\ell}\int^{\infty}_0\,\mathrm{d}\omega\,\Bigl(a^{\text{up}}_{\omega\ell m}u^{\text{up}}_{\omega\ell m}+a^{\text{in}}_{\omega\ell m}u^{\text{in}}_{\omega\ell m}\Bigr)+\text{h.c.}\,,
\end{aligned}
\label{vacBoulware}
\end{equation}
where $a^{\text{(in, up)}}_{\omega\ell m}$ are the respective annihilation operators, defining the Boulware state $  a^{\text{(in, up)}}_{\omega\ell m}|0_B\rangle =0$, and
 \begin{equation}
\begin{aligned}
u^{\text{(in, up)}}_{\omega\ell m}(\x)&=\frac{1}{\sqrt{4\pi\omega}}\Phi^{\text{(in, up)}}_{\omega\ell}(r)Y_{\ell m}(\theta,\phi)\expo^{-i\omega t}
\label{eq:4DSchw:BoulwareModes}
\end{aligned}
\end{equation}
are the respective field bases.   $\Phin$ and $\Phup$ are the normalized field modes obtained from Eq.~(\ref{norm}) with the normalization factor from Eq.~(\ref{Tin_norm}) and Eq.~(\ref{Tup_norm}). 

Therefore, for a static detector at a coordinate distance $R$ away from the black hole,
the Wightman function is\footnote{For a static detector, $r=r'=R$, $t-t^\prime = \Delta t=\Delta\tau/\sqrt{1-2M/R}$, and we can take $\theta=\theta'=\phi=\phi'=0$ without loss of generality.}
\begin{align}
W(\x,\x') &=\langle 0_B|\psi(\x)\psi(\x')|0_B\rangle \nonumber\\
&=\sum^{\infty}_{\ell=0}\int^{\infty}_0\,\mathrm{d}\omega\,\frac{\left(2\ell+1\right)}{16\pi^2\omega}\expo^{-i\omega\Delta\tau/\sqrt{1-2M/R}}\left(|\Phup(R)|^2+|\Phin(R)|^2\right)\,,
\label{W_BH_boulware}
\end{align}
with the field in the Boulware state. 
Substituting  \eqref{W_BH_boulware}  into the detector response rate~\eqref{eq:transRate} and commuting the $s$- and $\omega$-integrals, we obtain
\begin{equation}
\begin{aligned}
\mathcal{\dot{F}}_{\rm Boulware}\left(\Omega\right)=\frac{\Theta(-\Omega)}{8\pi|\Omega|}\sum^{\infty}_{l=0}\left(2\ell+1\right)\left(|\Phi^{\text{up}}_{\tilde{\omega}\ell}(R)|^2+|\Phi^{\text{in}}_{\tilde{\omega}\ell}(R)|^2\right)\,,
\label{Fdot_BH_Boulware}
\end{aligned}
\end{equation}
where $\tilde{\omega}:=\Omega\sqrt{1-2M/R}$.  Note that for the Boulware state, the response rate vanishes for positive energies, meaning that the detector can only de-excite, or alternatively $\Omega$ can only take negative values. In the following study we choose the variable $\Omega/T_{\rm loc}=8\pi M \tilde{\omega}$ as the effective gap parameter, where $T_{\text{loc}}$ is the local Hawking temperature, $T_{\text{loc}} = 1/(8 \pi M \sqrt{1-2M/R})$.

The other types of quantum vacuum fields for Schwarzchild spacetime, namely Hartle-Hawking state and Unruh state, have regular  stress-energy  across the future horizon (with Hartle-Hawking regular across the past horizon as well). 
As we explicitly show in Appendix~\ref{sec_HH}, the de-excitation rate of a static UDW detector with the field in the Hartle-Hawking (HH) and Unruh vacuum states overlap with that of the Boulware vacuum for energy gap $\Omega/T_{\rm loc}\lesssim -5$, and only deviates to a small finite value at zero gap limit ($ \mathcal{\dot{F}}_{\rm HH}\vert_{\Omega/T_{\rm loc}\to 0}\lesssim 0.01$), as shown in Figure~\ref{Fdot_HHtoB}b. Thus, we focus on the Boulware vacuum case in our following discussion.

\subsection{ECO}
An ECO has no event horizon and thus the vacuum must be of the Boulware type. Since only the in-modes satisfy the boundary condition at $r^*_0$, we must drop the up-mode contribution from the discussion above, yielding
\begin{equation}
\begin{aligned}
\psi=\sum^{\infty}_{\ell=0}\sum^{+\ell}_{m=-\ell}\int^{\infty}_0\,\mathrm{d}\omega\,\Bigl(a^{\text{in}}_{\omega\ell m}u^{\text{in}}_{\omega\ell m}\Bigr)+\text{h.c.}\,,
\end{aligned}
\end{equation}
instead of Eq.~(\ref{vacBoulware}), where 
\be
\begin{aligned}
u^{\text{in}}_{\omega\ell m}(\x)&=\frac{1}{\sqrt{4\pi\omega}}\Phi^{\text{in}}_{\omega\ell}(r)Y_{\ell m}(\theta,\phi)\expo^{-i\omega t}\,,\\
\end{aligned}
\end{equation}
with 
\be
\Phin=T_{\omega l}^\textrm{in}\frac{\rho_{\omega l}^\textrm{in}}{r}.
\label{Phin_ECO}
\ee
$\rho_{\omega l}^\textrm{in}$ is the in mode satisfying Eq.~(\ref{eom:rho}) with the boundary conditions parametrized by the reflectivity $R^{W}$ of the ECO~\cite{Conklin:2017lwb}
\begin{eqnarray}
\Phi_{\omega l}^\textrm{in}r&\to&\left\{\begin{array}{cc}
e^{-i\omega r^*_0}\,T_{\omega l}^\textrm{in}\left(e^{-i\omega(r^*-r^*_0)}+R^{\rm W}\,e^{i\omega(r^*-r^*_0)}\right), & r^*\to r^*_0  \\ 
e^{-i\omega r^*}+R_{\omega l}^\textrm{in}e^{i\omega r^*} & r^*\to\infty\end{array}\right.\label{rin_boundary}
\label{Phin_asym}
\end{eqnarray}
in contrast to \eqref{Phi_in_Asymtotic},
where $R^{\rm W}=1$ ($R^{\rm W}=-1$) corresponds to the Neumann (Dirichlet) boundary condition, whereas $R^{W}=0$ means the ECO is perfectly absorbing. From the solution $\rin$ to Eq.~(\ref{eom:rho}) satisfying the appropriate boundary condition at $r^*_0$, $\Tin$ and $\Rin$ are obtained by insisting on the correct behaviour at large $r^*$. We obtain
\be
\Tin=\left. \frac{2 \omega  e^{-i \omega r^*  }}{\omega \rin +i  \rin'} \right \vert_{r^*\to\infty},\quad \Rin= \left.\frac{  e^{-2i \omega r^* }(\omega \rin -i  \rin')}{\omega \rin +i  \rin'} \right \vert_{r^*\to\infty},
\label{TinRinNum}
\ee
where in practice $\rin$ and its derivatives $\rin'$ are evaluated at a sufficiently large $r^*$. As shown in Appendix C, we can introduce another set of ECO solutions similar to~(\ref{Phi_up_Asymtotic}) as an analog of the up-mode $\rho_{\rm up}$ in the black hole setting, but only for mathematical use since only the in-modes satisfy the boundary condition at $r^*_0$. In this way, it is straightforward to show that the transmission and reflection coefficients of the ECO can be cast into the same form as Eq.~(\ref{Tin_norm}).  Energy conservation implies 
 \be
 |\Tin|^2 (1-(R^{W})^2)=1-|\Rin|^2,
 \label{AbsTinRin}
 \ee
so that $|\Rin|=1$ for $R^W=\pm 1$.

Then, with the normalized $\Phin$ field, we find
\begin{equation}
W_{\rm ECO}(\x,\x')=\sum^{\infty}_{\ell=0}\int^{\infty}_0\,\mathrm{d}\omega\,\frac{\left(2\ell+1\right)}{16\pi^2\omega}\expo^{-i\omega\Delta\tau/\sqrt{1-2M/R}}|\Phin(R)|^2\,,
\label{eq:4DSchw:BoulwareStaticW}
\end{equation}
for the Wightman function, and in turn
\begin{equation}
\begin{aligned}
\mathcal{\dot{F}_{\rm ECO}}\left(\Omega\right)=\frac{\Theta(-\Omega)}{8\pi|\Omega|}\sum^{\infty}_{l=0}\left(2\ell+1\right)|\Phi^{\text{in}}_{\tilde{\omega}\ell}(R)|^2\,,
\label{Fdot_ECO_Boulware}
\end{aligned}
\end{equation}
for the response rate.
\section{Results}
\label{result}

\begin{figure}[h]
 \centering
    \includegraphics[width=0.49\linewidth]{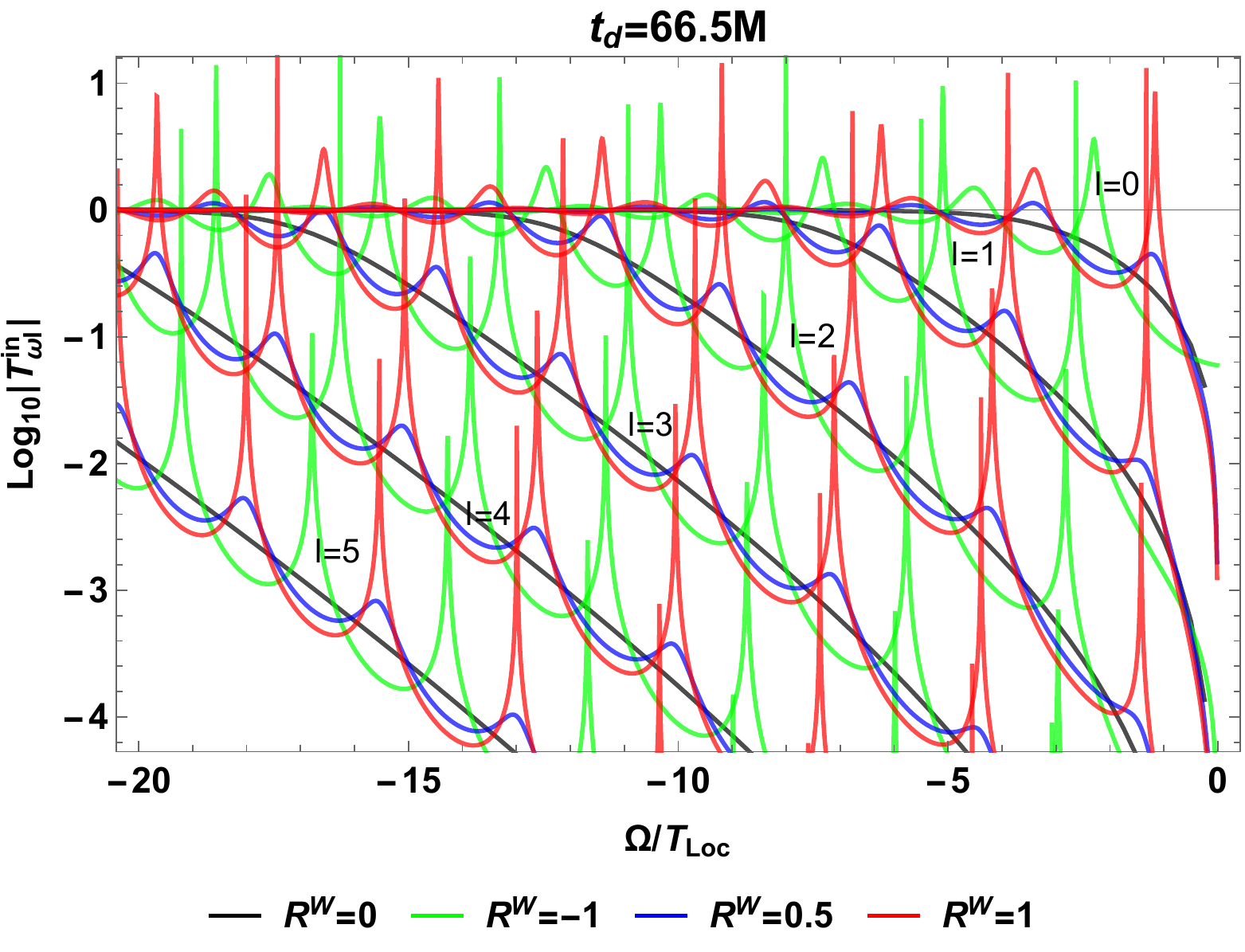}  
\includegraphics[width=0.49\linewidth]{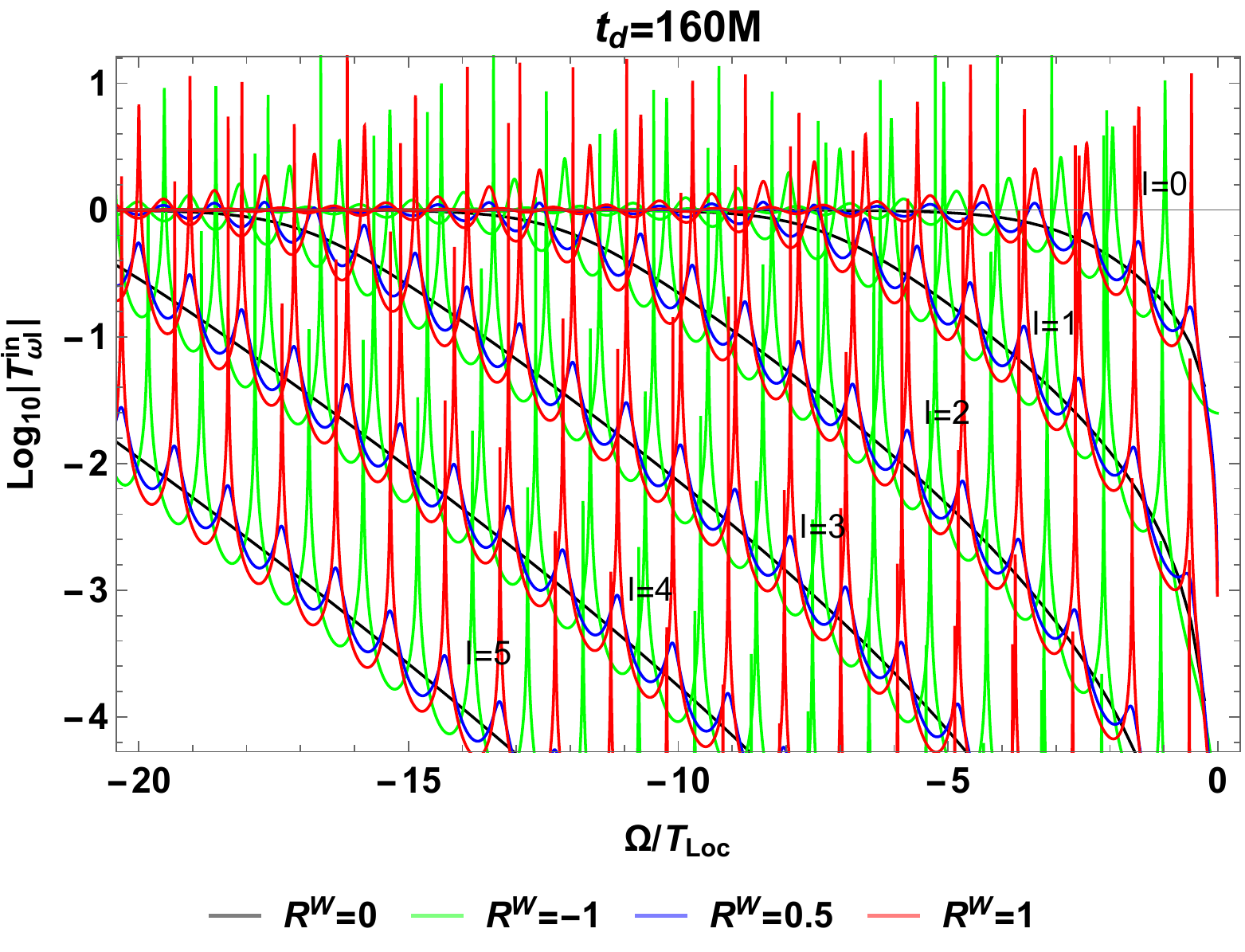}  
\caption{$\log_{10} |\Tin|$ for (a) small ($t_d= 66.5 M$) and (b) large ($t_d=160M$) time delays with $R^{W}=(0,-1,0.5,1)$ for $l=0\sim5$.}
    \label{TinLog}
\end{figure}

The most distinctive feature in the comparison of an ECO to a black hole is the resonance spectrum of $\Tin$ (i.e., the transfer function). In Figure~\ref{TinLog}, we show the $\Tin$ spectrum of two benchmark models with small $(t_d=66.5\,M)$ and large $(t_d=160M)$ time delays.
The position and width of these resonances are determined from the real and imaginary parts of the complex poles of $\Tin$, which are   affected by both the ECO boundary condition and the gravitational potential. The spacings of the resonance spikes is related to the time delay such that\footnote{Note that the actual $\Delta \omega$ has a small frequency dependence resulting from the finite width of the gravitational potential around $r^*_{\rm peak}$.}
\be
\Delta{\omega}= \frac{2\pi}{t_d}, \quad  \Delta{\Omega/T_{\rm loc}}=8\pi M \Delta{\tilde{\omega}}= \frac{16\pi^2M}{t_d}.
\label{Tx}
\ee
Therefore, the resonance spacing is smaller for a larger time delay (i.e., when $r_0$ is closer to the would-be horizon). For a zero $R^W$, $\Tin$ has no resonances at all since the boundary is perfectly absorbing, and thus is equivalent to the $\Tin$ of a black hole. Differing time delays in this case do not affect the result since different locations of a perfect-absorbing boundary are equivalent for the wave modes. As we will demonstrate, these resonance structures will also show up in  the UDW detector's response rate, yielding distinct structures in contrast to the black hole.

\begin{figure}[h]
 \centering
    \includegraphics[width=0.48\linewidth]{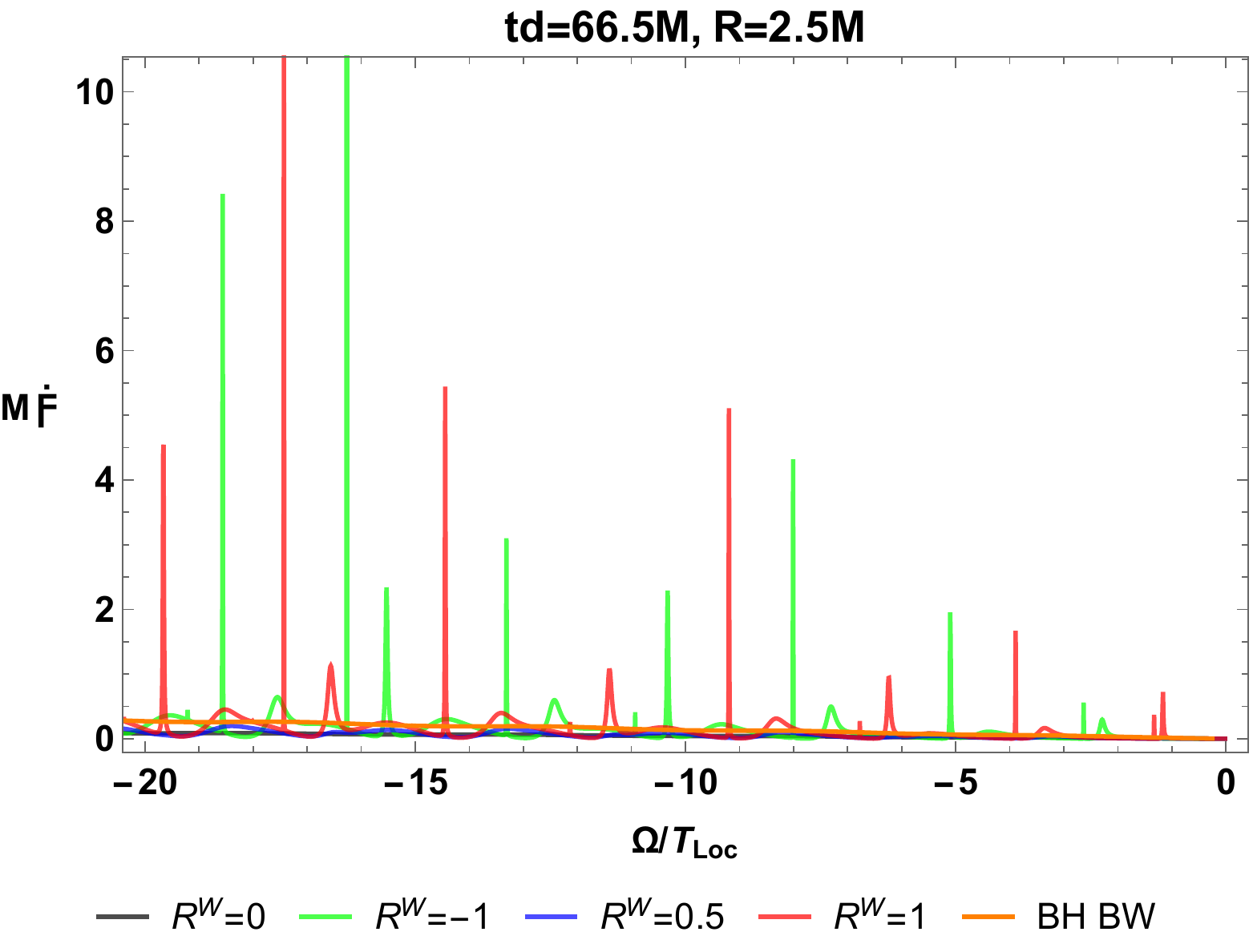}  
\includegraphics[width=0.48\linewidth]{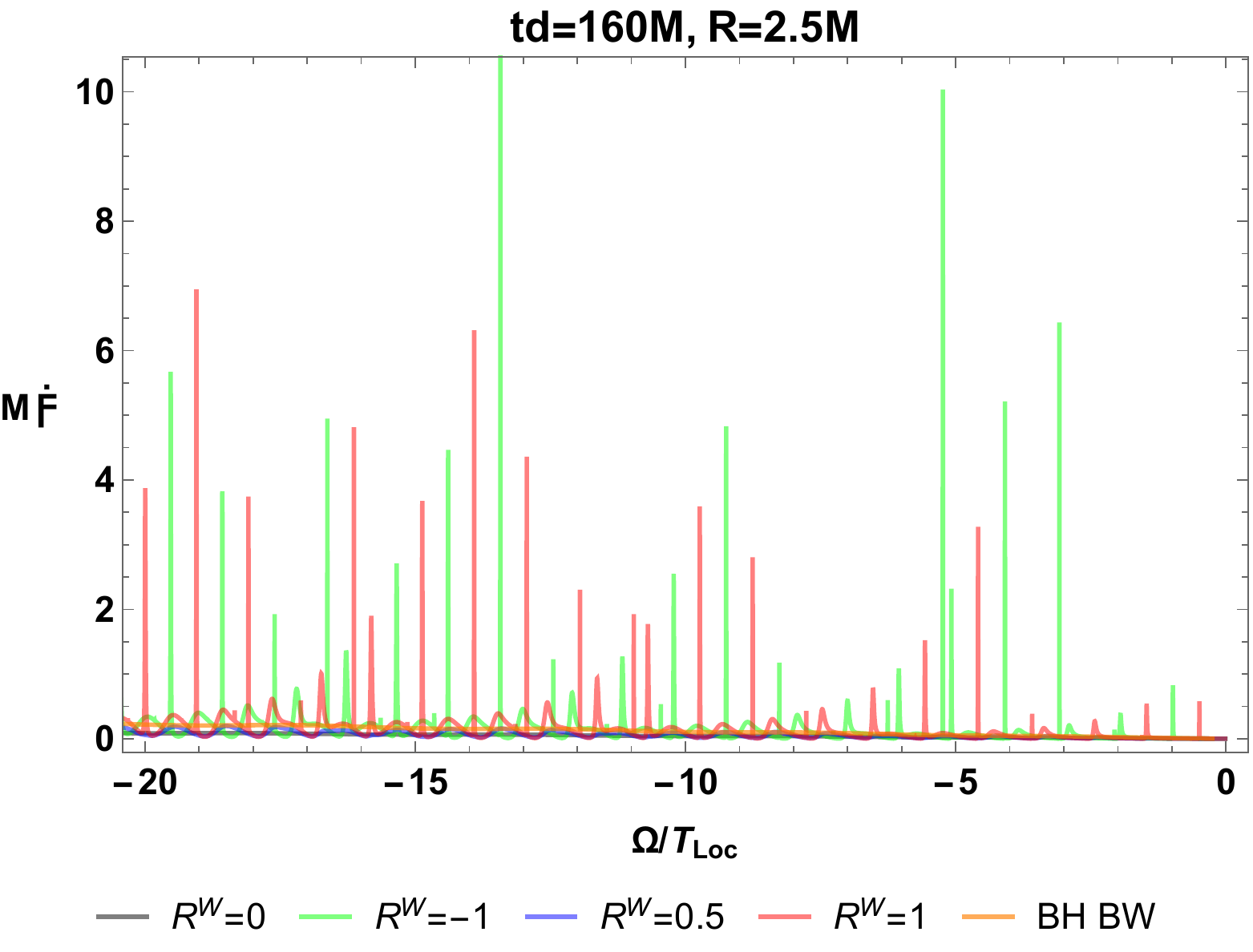}  
    \includegraphics[width=0.48\linewidth]{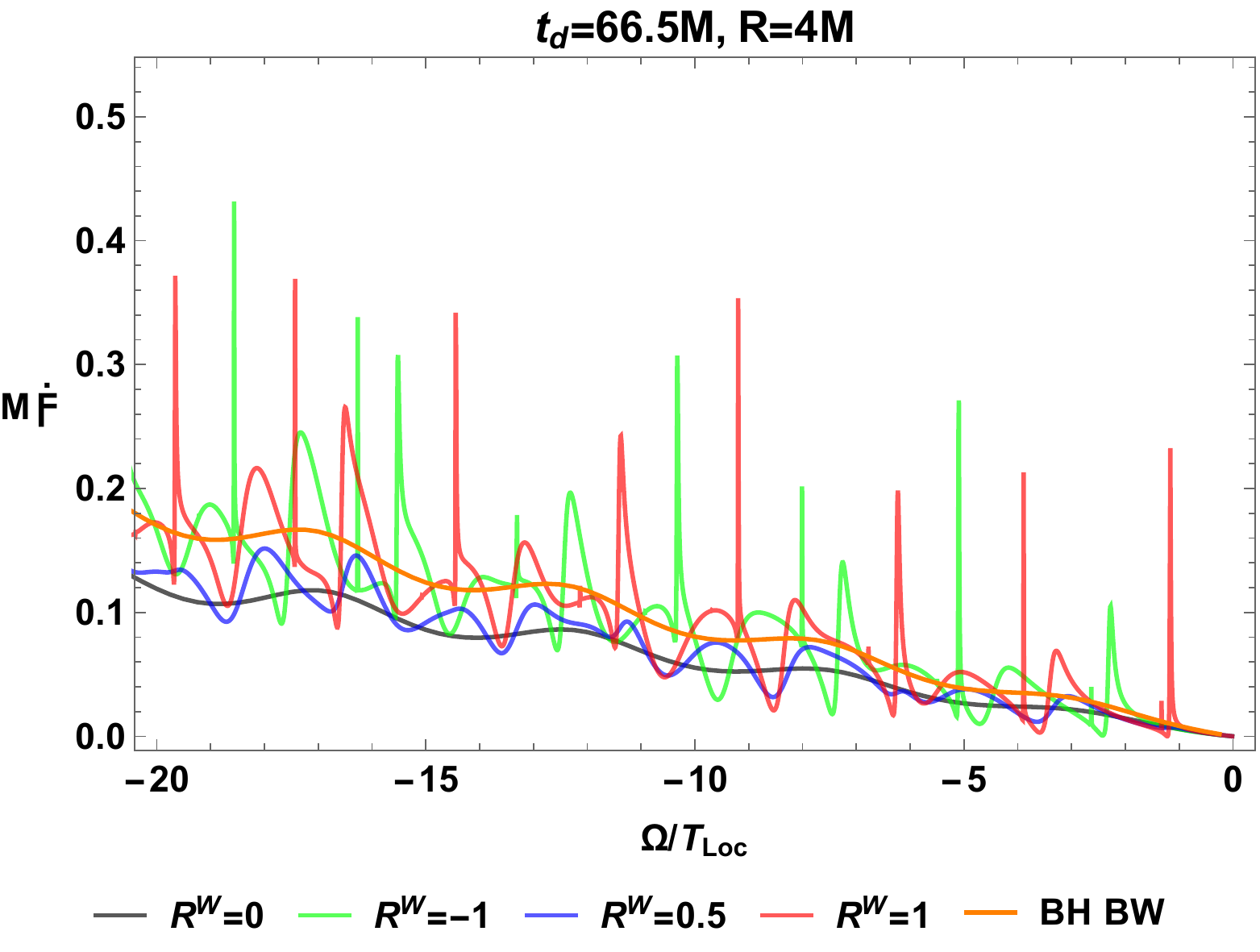}  
\includegraphics[width=0.48\linewidth]{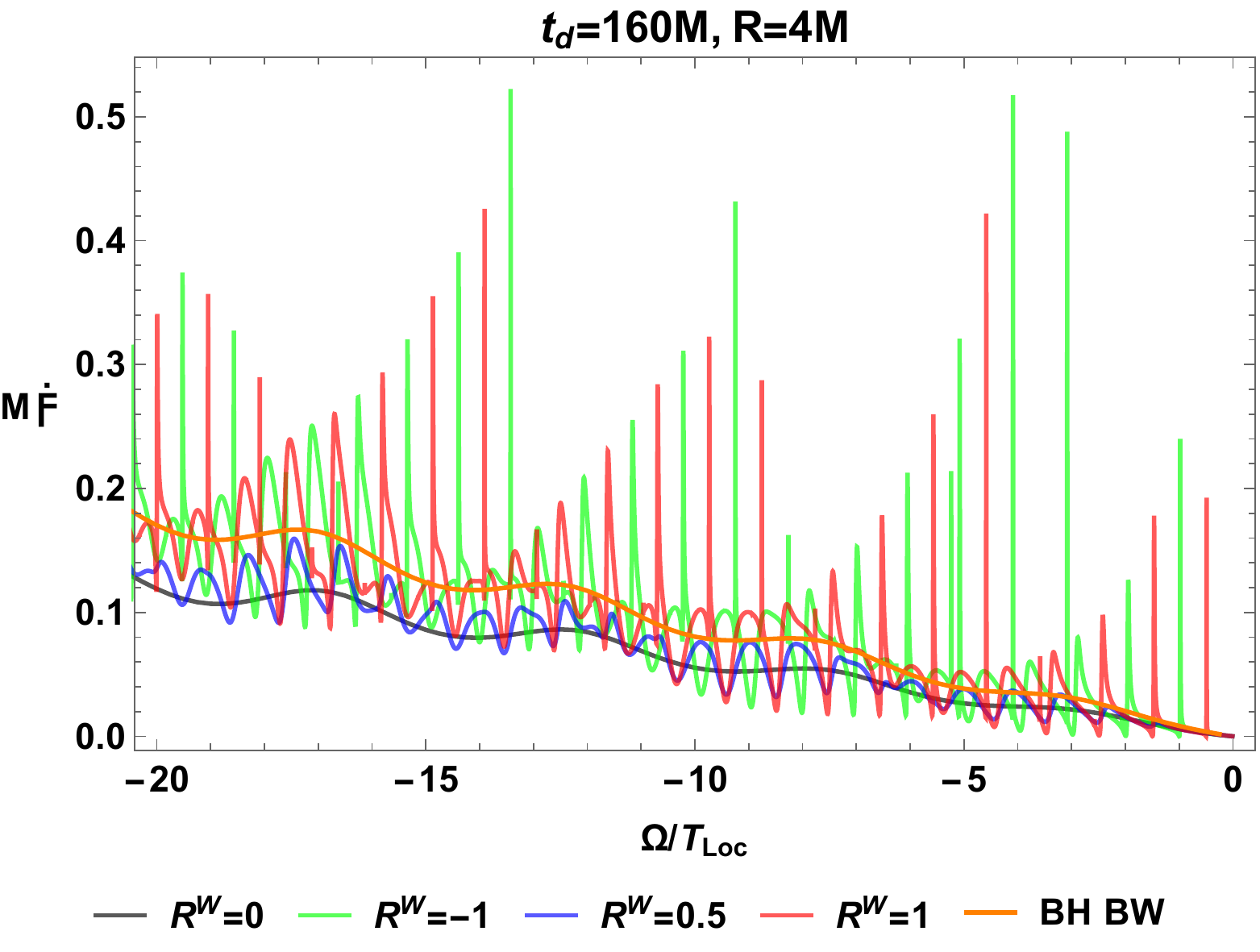}  
\includegraphics[width=0.48\linewidth]{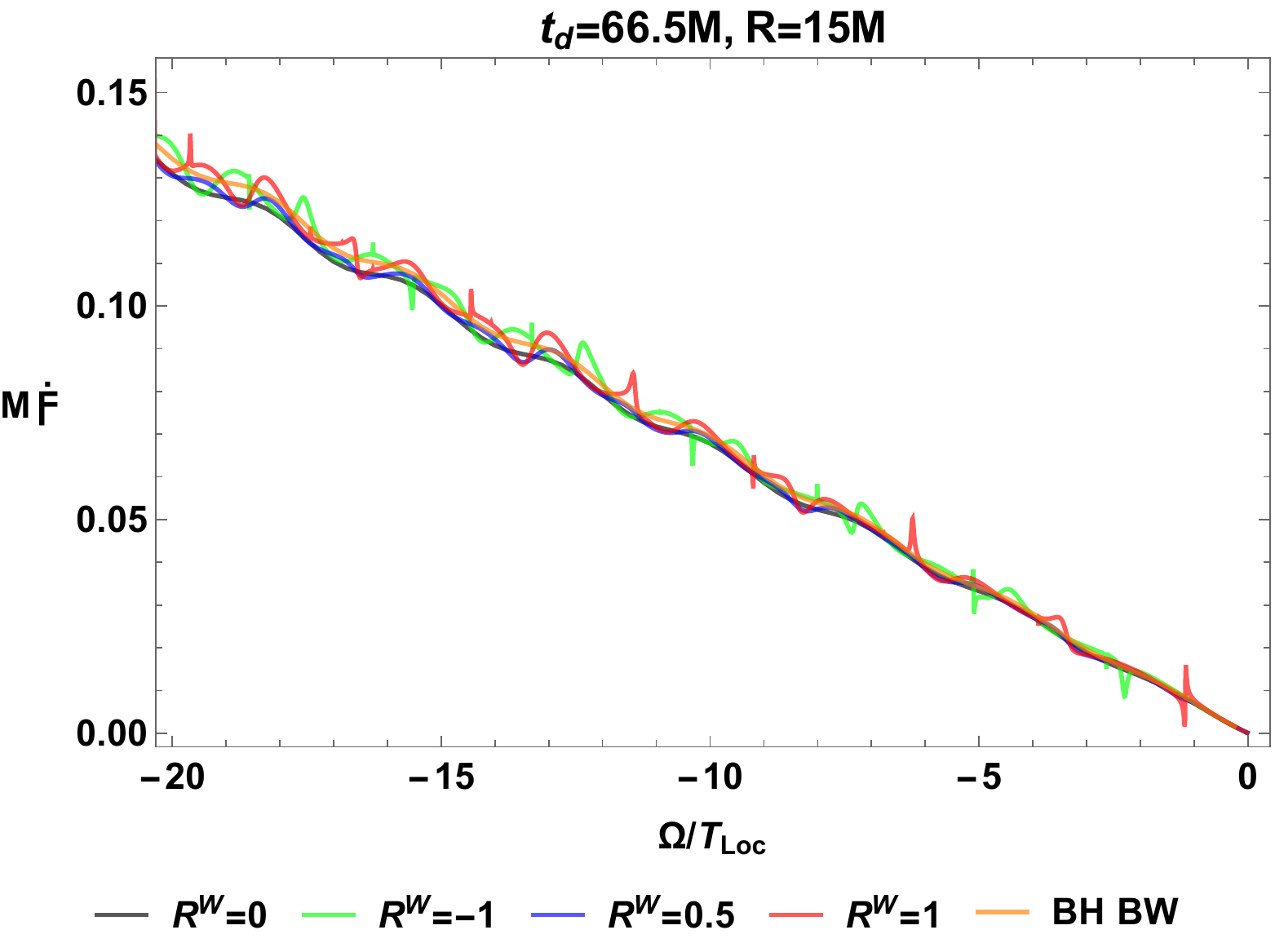}  
    \includegraphics[width=0.48\linewidth]{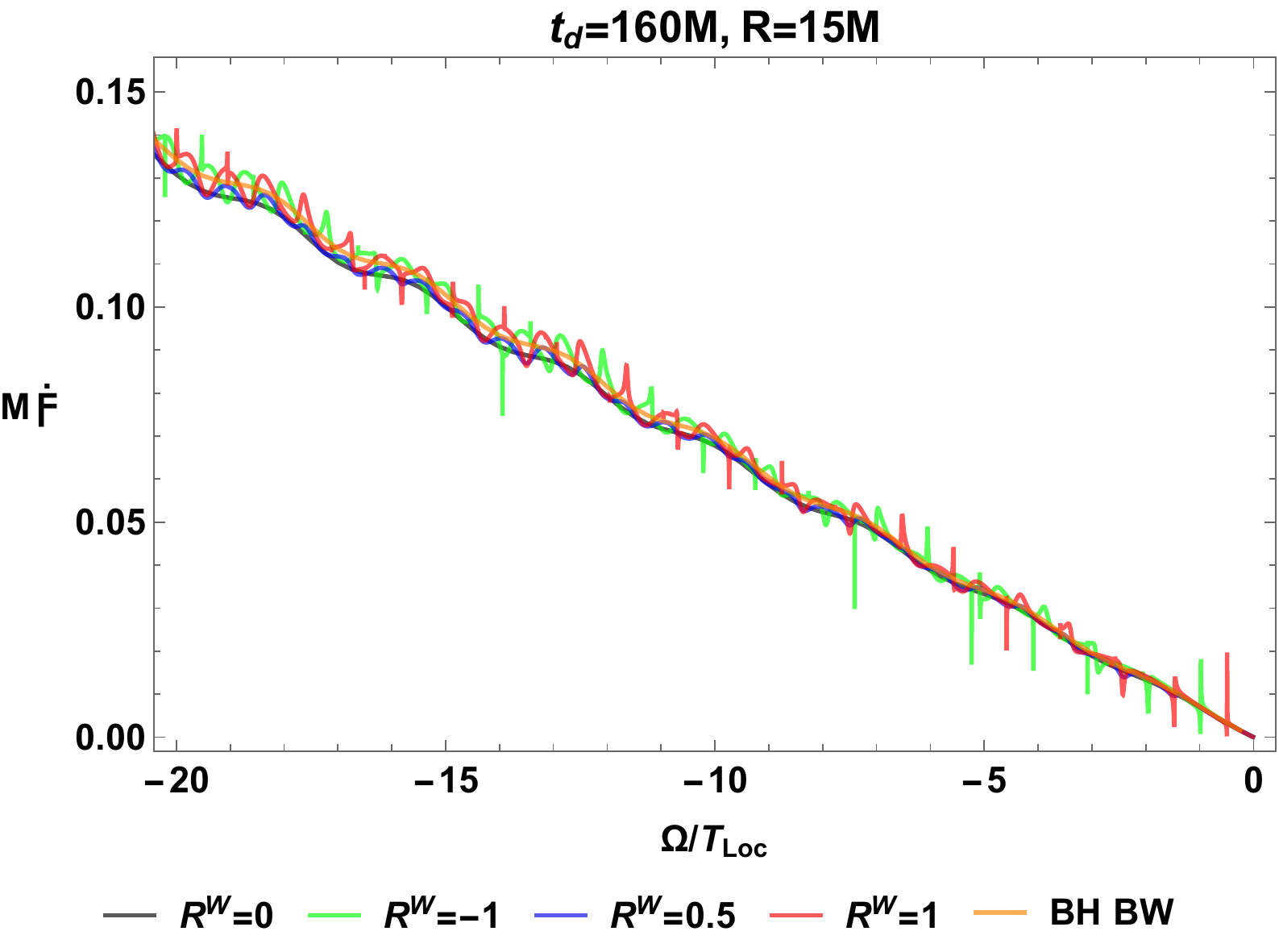}  
 \caption{$M\dot{\mathcal{F}}$ as a function of $\Omega/T_{\text{loc}}$ for a static detector at $ R=2.5M, 4M, 15M$ (top to bottom) above an ECO of (left) small ($t_d= 66.5\,M$), and (right) large ($t_d=160M$) time delays with $R^{W}=(0,-1,0.5,1)$. The orange line is for the black hole case with Boulware vacuum.}
    \label{Fdot_all}
\end{figure}

Using~(\ref{Fdot_ECO_Boulware}),  the response rate can be calculated with the $l$ sum truncated to $l_{\rm max}$ such that the contribution with $l=l_{\rm max}+1$ is negligible.  It turns out that $l_{\rm max}$ increases for a larger radial distance $R$ and a larger $|\Omega/T_{\rm loc}|$, and is insensitive to the boundary condition and time delay changes. For example, when $|\Omega/T_{\rm loc}|$ goes to 20, $l_{\rm max}\approx 5$ for $R=4M$, and $l_{\rm max}\approx 15$ for $R=15M$.  We give an analytic approximation of $l_{\rm max}$ in Appendix~\ref{sec_lmax}. 
To illustrate some general features, we present in Figures~\ref{Fdot_all} the response rate for the previous two benchmark examples with different  ($t_d$,$R$).

Note that the first row of Figure~\ref{Fdot_all} is for the case where the static detector is inside the cavity ($R=2.5M$), in contrast to the second and third rows where the detector is placed outside ($R=4M$ and $R=15M$).

We easily see the striking resonance patterns of the response rate in the first row of Figure~\ref{Fdot_all}, which match those of $|\Tin|$ shown in Figure~\ref{TinLog} that are independent of $R$; cases with the smaller time delay have resonances more sparse in distribution, and vice versa.   
Comparing to those of the second and third rows, we see as the distance gets larger, the resonances become more suppressed in size. We give a detailed analysis in the next section.

In the second and thirds rows of Figure~\ref{Fdot_all}, we see a general trend for the detector’s response rate to grow roughly linearly with $|\Omega/T_{\rm loc}|$. But in the second row, we see that an ECO with $R^W = 0$ has a response rate that is relatively lower than that of the black hole, even though the boundaries of both objects are perfectly absorbing. Comparing Eq. (18) and Eq. (26), we see that this is because the ECO case has no up-mode contribution for any $R^W$. By comparison to the second row, the third row then shows that the extra up-mode contribution in the black hole response becomes relatively smaller for increasing $R$. In the large $R$ limit all the response rates approach the Minkowski vacuum state result, which is proportional to $-\Theta(-\Omega)\Omega$.
\section{Analysis}
\begin{figure}[h]
 \centering
        \includegraphics[width=0.48\linewidth]{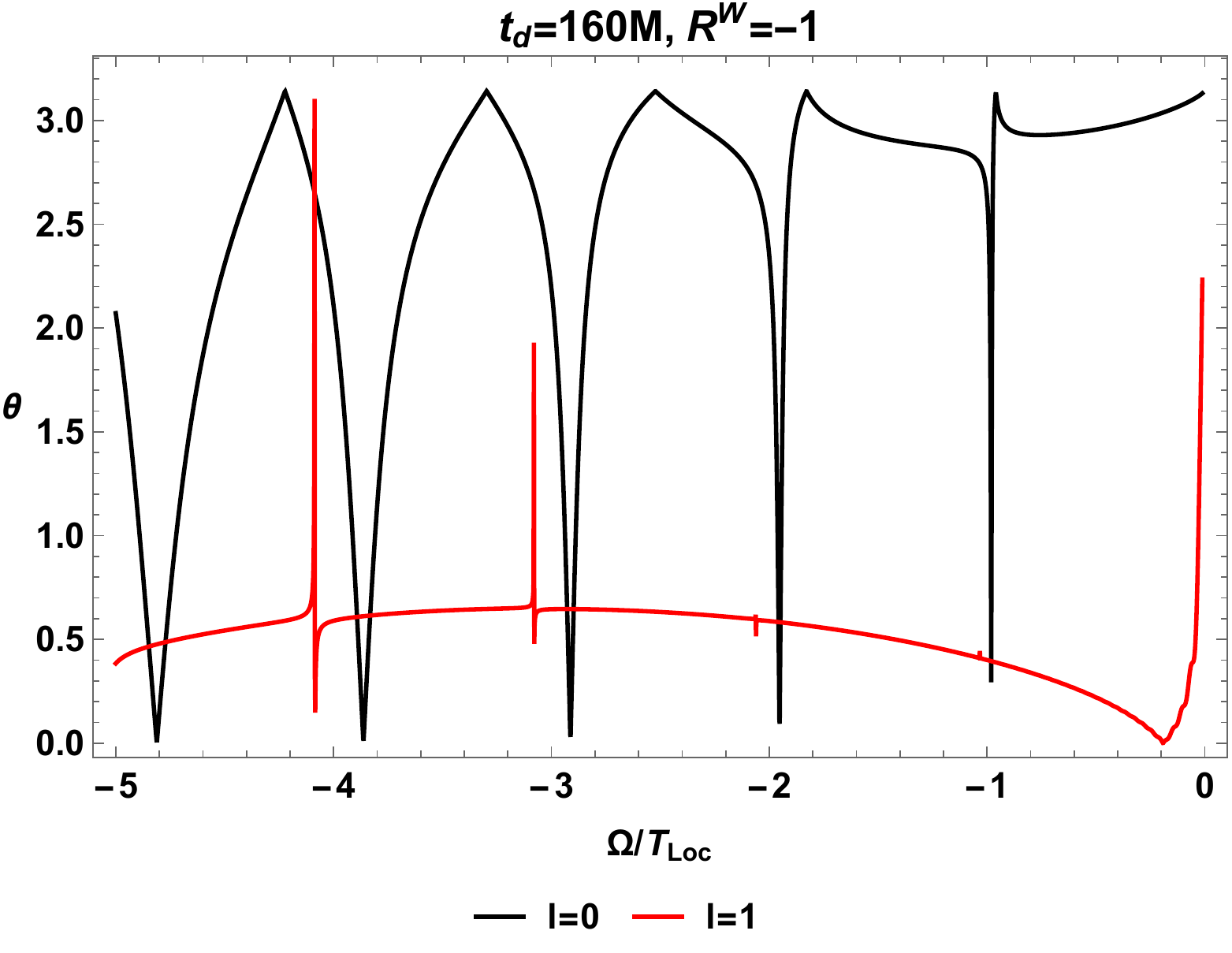}  
     \includegraphics[width=0.512\linewidth]{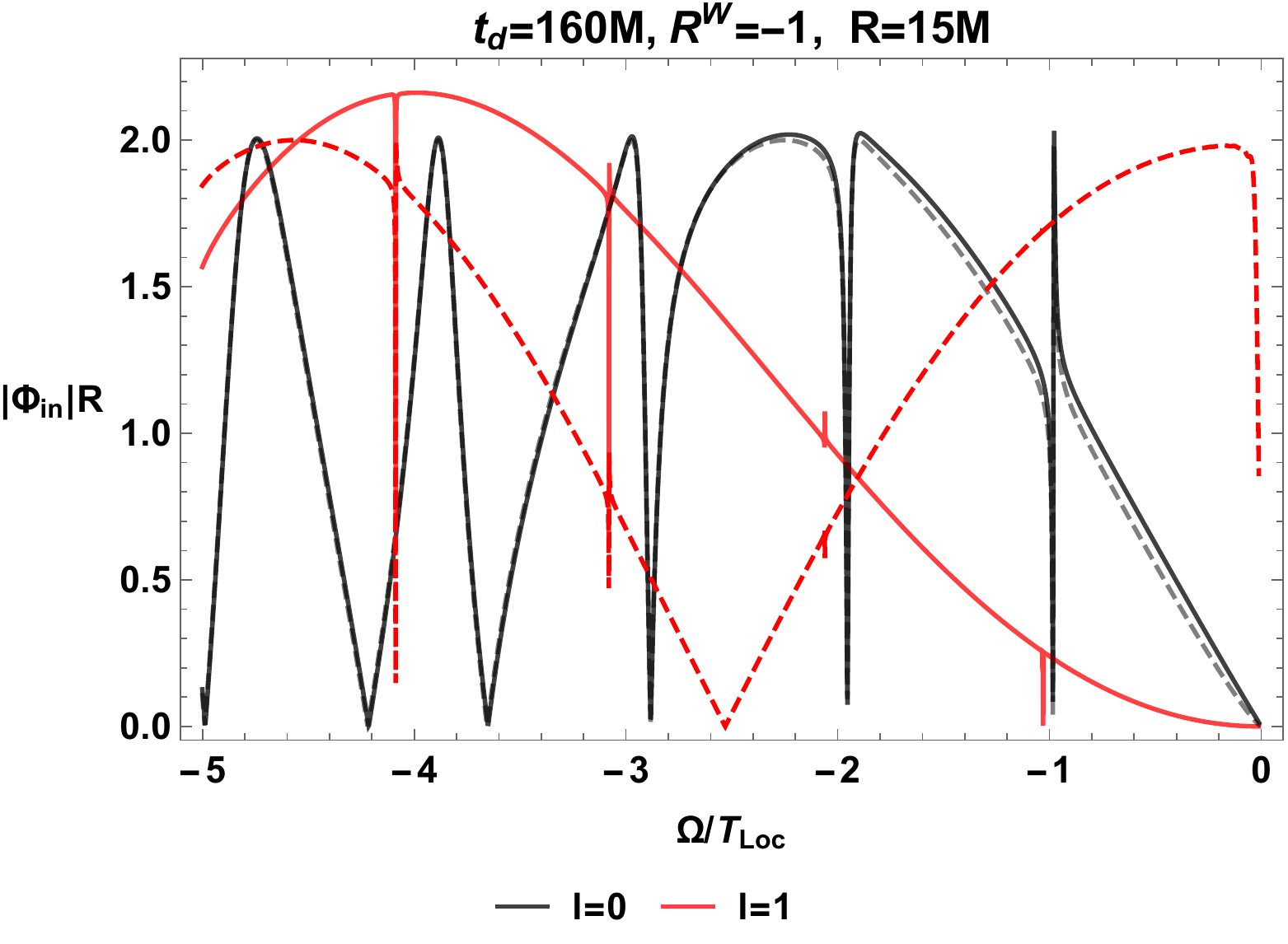}  
 \caption{(a)  $\theta=\arccos(\Re(\Rin)/|\Rin|)$  and (b) $|\Phin| R$ ($R=15M$)   for a time delay $t_d=160M$ with $R^W=-1$ for $l=0$ (black) and $l=1$ (red). Solid lines are the exact results from Eq.~(\ref{Phin_ECO}), while dashed lines denote the approximation from Eq.~(\ref{Phin_bigr}). }
    \label{theta_Phi_td160M_RWn1}
\end{figure}

From Eq.~(\ref{Phin_asym}), we see the behaviour of $\Phin$ and the resulting $\mathcal{\dot{F}_{\rm ECO}}$ is dominated by $\Tin$ at small $r$ and by $\Rin$ at large $r$ up to some complex phase factors. To see this more explicitly, we can take absolute value of $\Phi_{\omega l}^\textrm{in}$, obtaining
\begin{align}
|\Phi_{\omega l}^\textrm{in}|r &\to |T_{\omega l}^\textrm{in}| \sqrt{1+R^{W^2} +2 R^W \cos [2\omega (r^*- r^*_0)]}, \text{ as } r^*\to r^*_0  
\label{Phin_smallr}  \\
|\Phi_{\omega l}^\textrm{in}|r &\to \sqrt{1+ |\Rin|^2+ 2|\Rin| \cos(2\omega r^*+\theta)} , \quad\,\,\,\, \text{  as  } r^*\to\infty.
\label{Phin_bigr}
\end{align}
from Eq.~(\ref{Phin_asym}), 
where $\theta=\arccos(\Re(\Rin)/|\Rin|)$.  The value of $\theta$  can also have large fluctuations at the location of  the $\Tin$ resonances, as the benchmark model in Figure~\ref{theta_Phi_td160M_RWn1}a shows. We see that  $|\Phin|r$ goes to $|T_{\omega l}^\textrm{in}| (1+R^W)$ when approaching  the ECO boundary,  and at large $r$ to some value between $(1-|\Rin|)$ and $(1+|\Rin|)$, as the example in Figure~\ref{theta_Phi_td160M_RWn1}b illustrates.  From Eq.~(\ref{AbsTinRin}), we see that $|\Rin| \leq 1$ for $R^W\leq 1$, but $|\Tin|$ can display relatively large resonance peaks. This, in conjunction with  (\ref{Phin_smallr}) and  (\ref{Phin_bigr}), partially explains why $|\Phin| r$ and the corresponding contribution in $\mathcal{\dot{F}}$ at given $l$ exhibit smaller resonances at a larger distance.

Note that Eqs.~(\ref{Phin_smallr}) and~(\ref{Phin_bigr}) give a good match to the exact result~(\ref{Phin_ECO}) for $|r^*-r^*_{\rm peak}|$ sufficiently large, that is provided $\omega^2$ is larger than $V(r)$, thus mapping to the parameter space of large $\omega$ and small $l$. These features can also be seen in the example shown in Fig~(\ref{theta_Phi_td160M_RWn1})b, where the exact result of $|\Phin|r$ at $l=0$ matches the approximate form well, with resonances fluctuating between 0 and +2, while all larger $l\geq1$ show notable deviation at small $|\Omega/T_{\rm loc}|$ $(=8\pi M \, |\omega|)$.

\begin{figure}[h]
 \centering
    \includegraphics[width=0.48\linewidth]{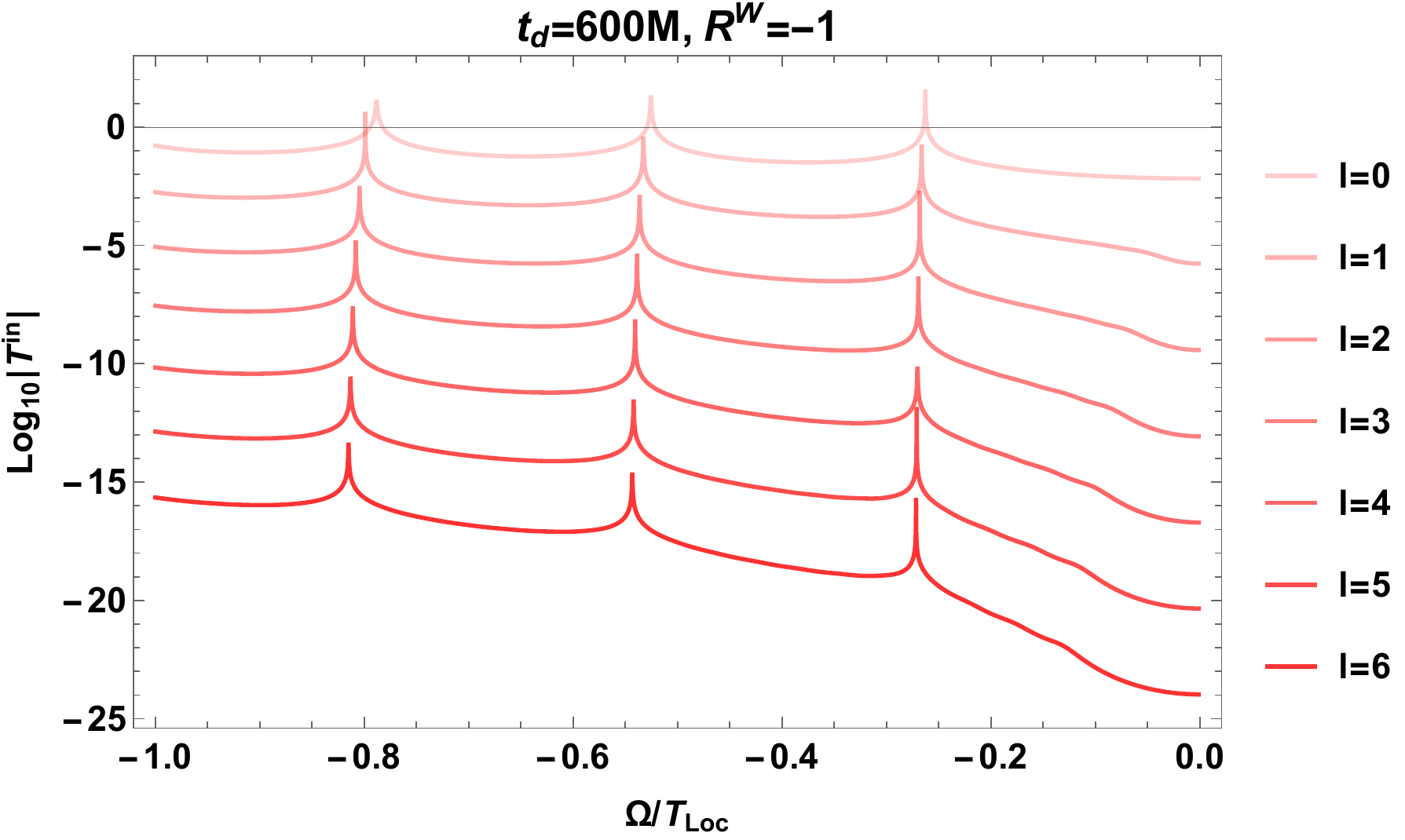}  
     \includegraphics[width=0.51\linewidth]{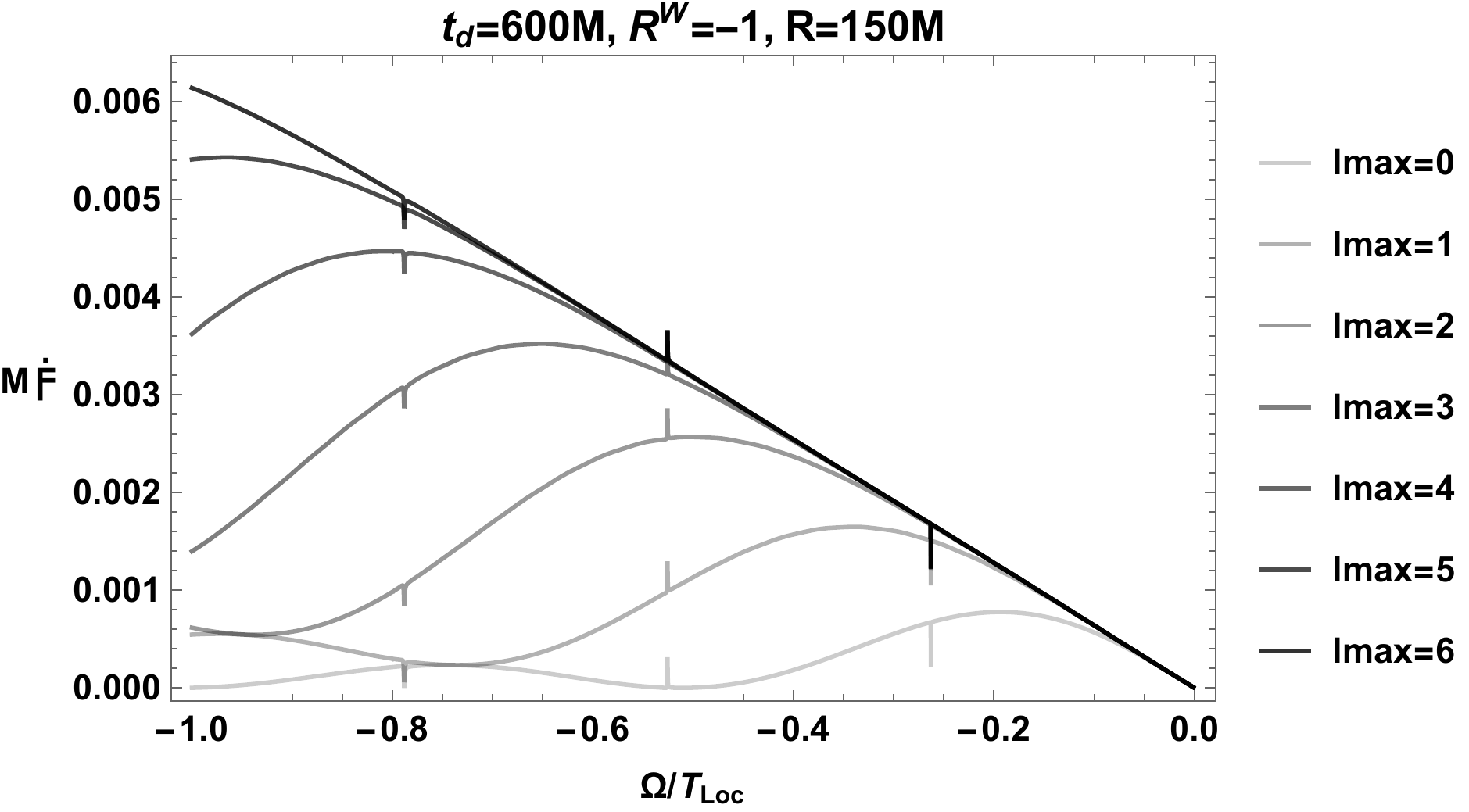}  
 \caption{(a) $\log_{10} |\Tin|$ and  (b) $M\dot{\mathcal{F}}$ for a large time delay $t_d=600M$ with $R^W=-1$. The $M\dot{\mathcal{F}}$ result has the static detector at a large distance $R=150M$ with $l$ summed from zero to different $l_{\rm max}$.}
    \label{td600M_R150M}
\end{figure}

As Figure~\ref{TinLog} shows, the resonance structures of $\Tin$ (and thus the similar structure of $\Rin$) shift to a larger $|\Omega/T_{\rm loc}|$  for a larger $l$. As $r$ increases, $l_{\rm max}$ also becomes large (Appendix~\ref{sec_lmax}), so that the $l$-summed results have more dependence on large $l$ yielding a milder resonance structure at a given frequency.  This also illustrates why the $\mathcal{\dot{F}}$ results in Figures~\ref{Fdot_all} show milder resonances at a larger distance for a given $\Omega/T_{\rm loc}$. This further implies that we can identify a clear resonance structure at   very large $R$ provided the time delay is large so that the resonances are located at small frequencies where  $l_{\rm max}$ is small. For illustration, we give a benchmark example of a very large time delay $(t_d=600M)$ with the detector at a very large distance $R=150M$ in Figure~\ref{td600M_R150M}. We can see that the resonances of $|\Tin|$ at small $l$ in Figure~\ref{td600M_R150M}a are clearly reflected in the response rate result shown in Figure~\ref{td600M_R150M}b, even at this large $R$.
\section{Case $R^W=-0.98$ }
\begin{figure}[h]
 \centering
\includegraphics[width=0.49\linewidth]{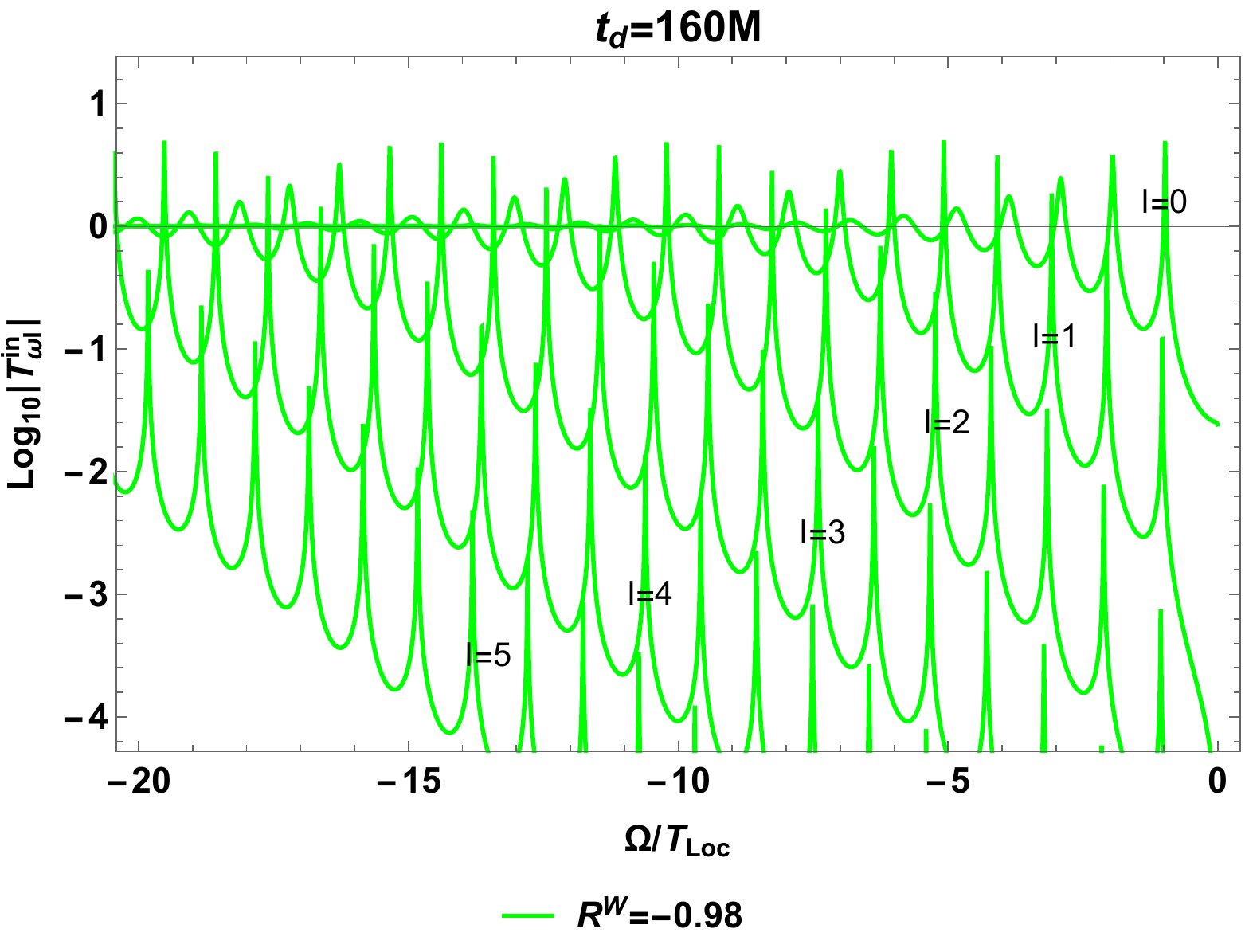}  
 \includegraphics[width=0.49\linewidth]{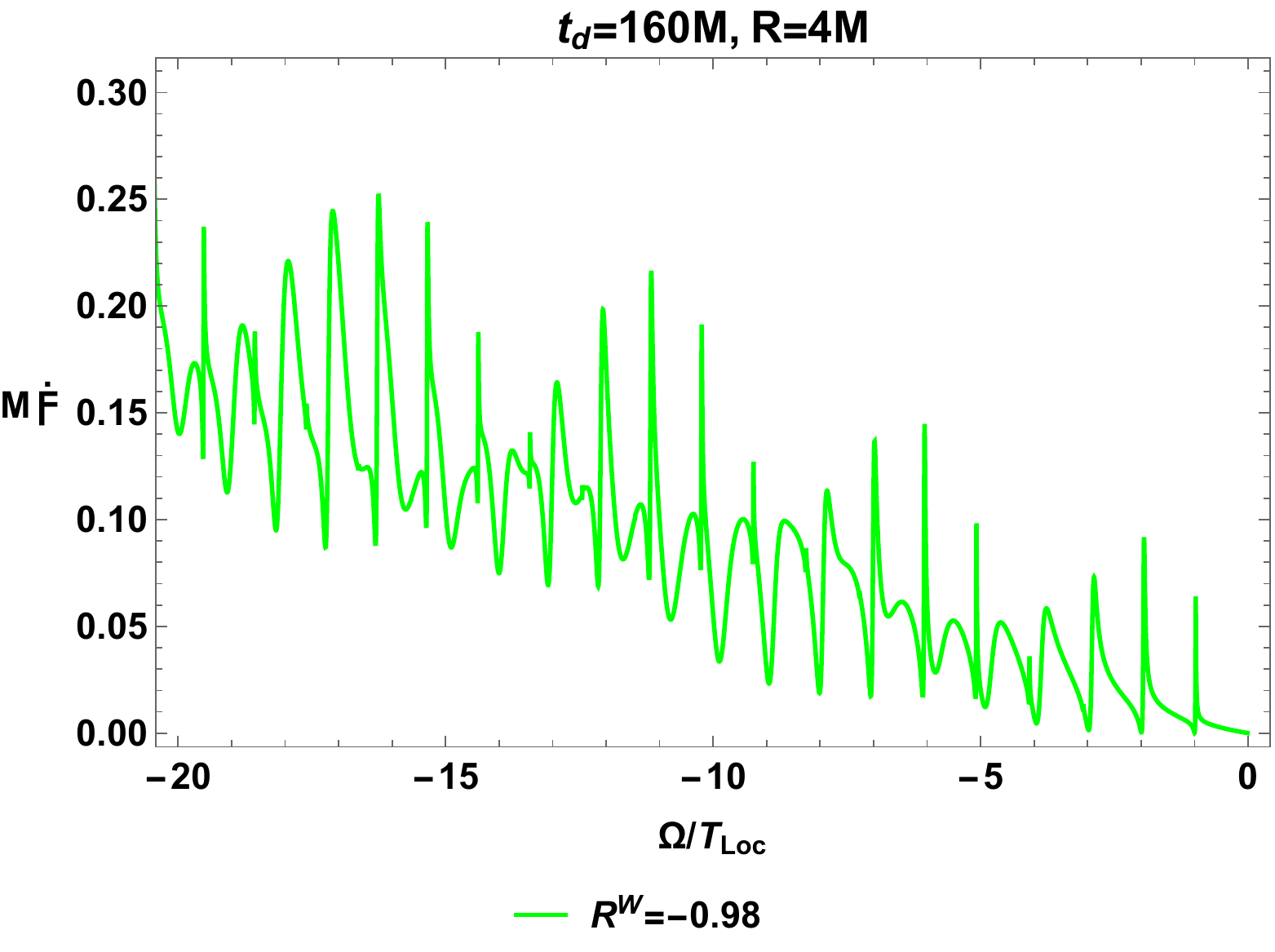}  
 \caption{(a) $\log_{10} |\Tin|$ and (b) $M\dot{\mathcal{F}}$ ( $R=4M$) as a function of $\Omega/T_{\text{loc}}$ for a time delay $t_d=160M$ with $R^{W}=-0.98$.}
    \label{Fdot_R4M_RW098}
\end{figure}

For physical realizations of the ECO model, $|R^W|$ can be taken to be slightly less than one to account for a small amount of damping. Damping may be needed in the case of spinning ECOs to avoid an instability. In the case of a static 2-2-hole the damping may be explicitly calculated, and the resulting frequency dependent damping is largest for small frequencies~\cite{Holdom:2020onl}. For our static ECO study we shall illustrate the effect of damping by choosing a constant value $R^W=-0.98$. We show in Figure~\ref{Fdot_R4M_RW098}a  the $\Tin$ function with large time delay $t_d=160M$, with  the related response rate  at small distance ($R=4M$)
depicted in Figure~\ref{Fdot_R4M_RW098}b. Comparing to the $R^W=-1$ case shown in Figure~\ref{TinLog}b and the right figure in the second row of Figure~\ref{Fdot_all}, we observe that the response rate for $R^W=-0.98$ is close to the $R^W=-1$ case, except that the resonance heights of $\Tin$ and $\mathcal{\dot{F}_{\rm ECO}}$ are both noticeably milder than those for $R^W=-1$. It turns out that this feature is also present in all other examples we have examined, including the large-$R$ cases.

\begin{figure}[h]
 \centering
    \includegraphics[width=0.49\linewidth]{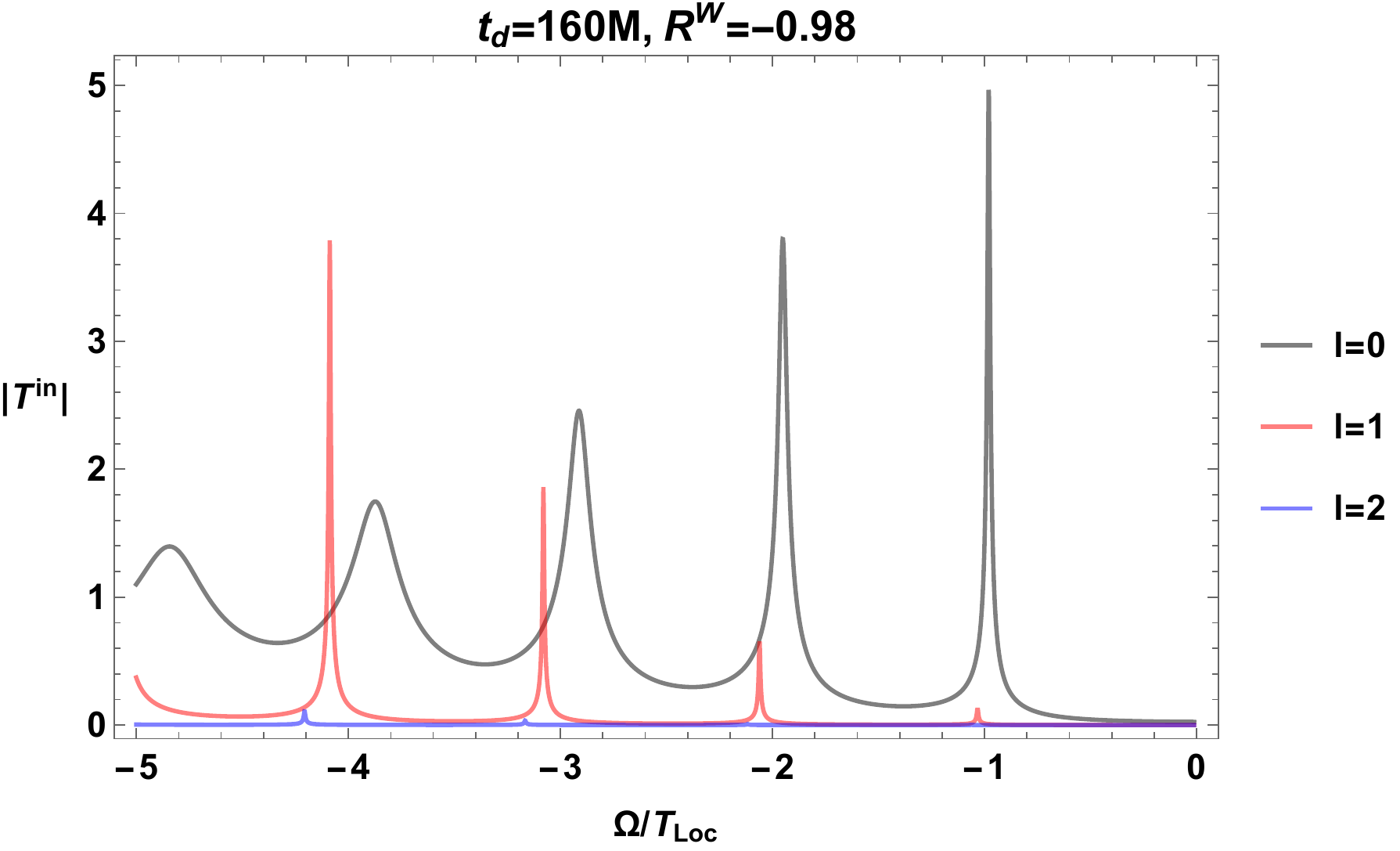}  
\includegraphics[width=0.49\linewidth]{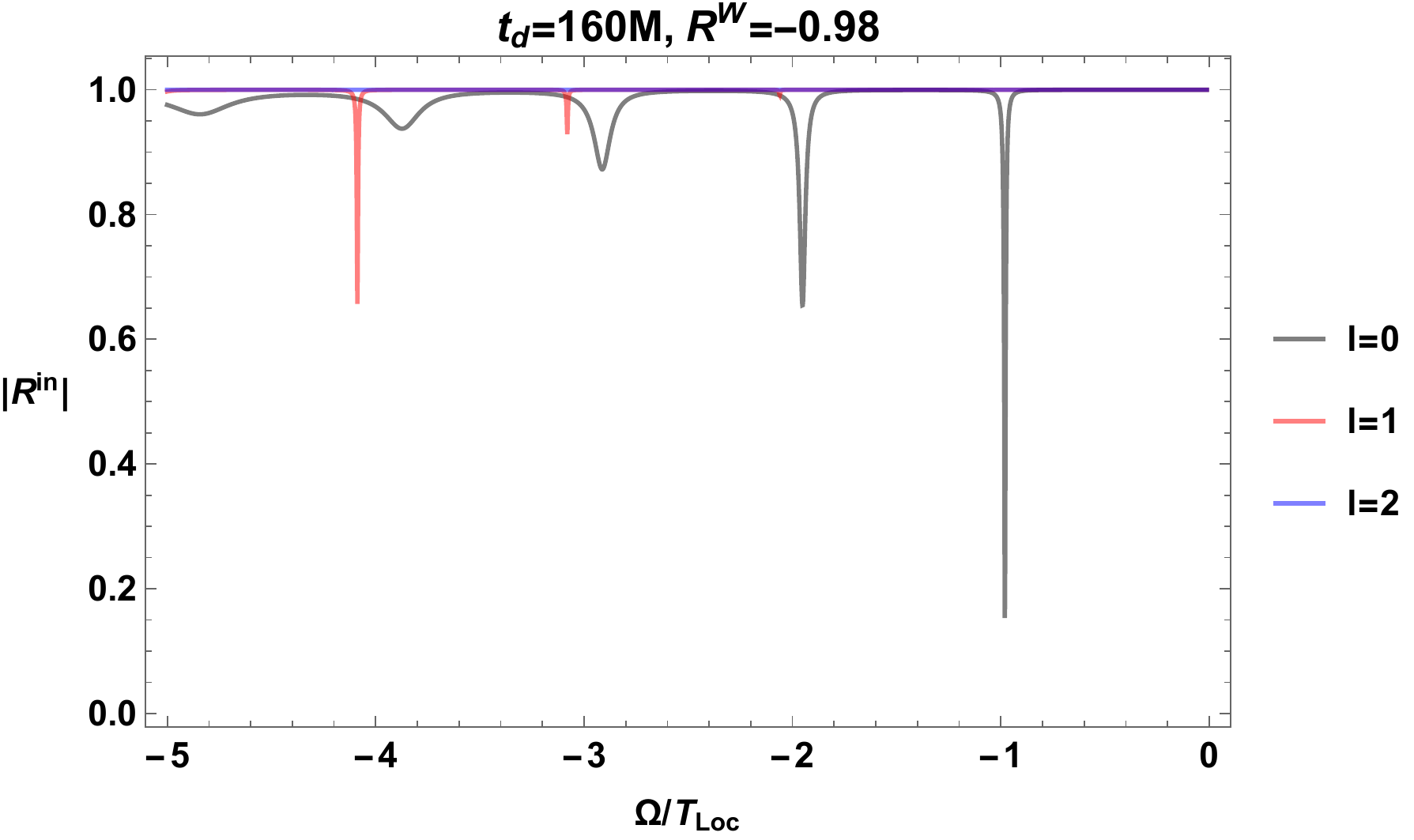}  
 \caption{(a) $ |\Tin|$ and (b)$ |\Rin|$ for a time delay $t_d=160M$ with $R^W=-0.98$.}
    \label{TinRin098}
\end{figure}

For the $R^W=-0.98$ examples at large $R$,  the exact result of $|\Phi_{\omega l}^\textrm{in}|r$ can be approximated by Eq.~(\ref{Phin_bigr}). From~(\ref{AbsTinRin}), when $|R^W|$ is close but not equal to unity, the resonances of $\Tin$ can show up as dips in $|\Rin|$, as shown in Figure~\ref{TinRin098} for a benchmark example with a time delay $t_d=160M$. These dips can potentially translate to large dips in $M\dot{\mathcal{F}}$  at  large $R$, considering Eq.~(\ref{Phin_bigr}).

\begin{figure}[h]
 \centering
         \includegraphics[width=0.473\linewidth]{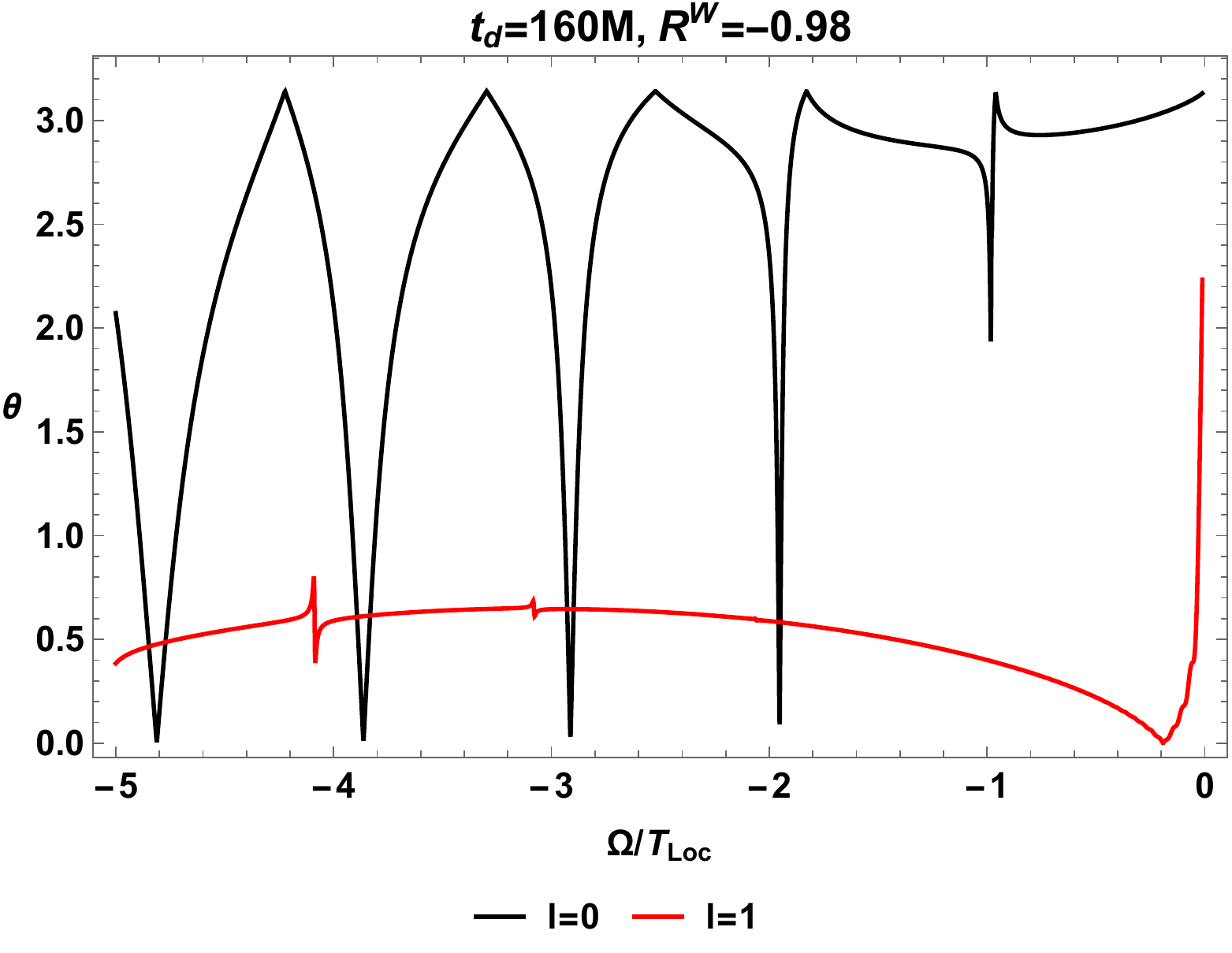}  
           \includegraphics[width=0.5\linewidth]{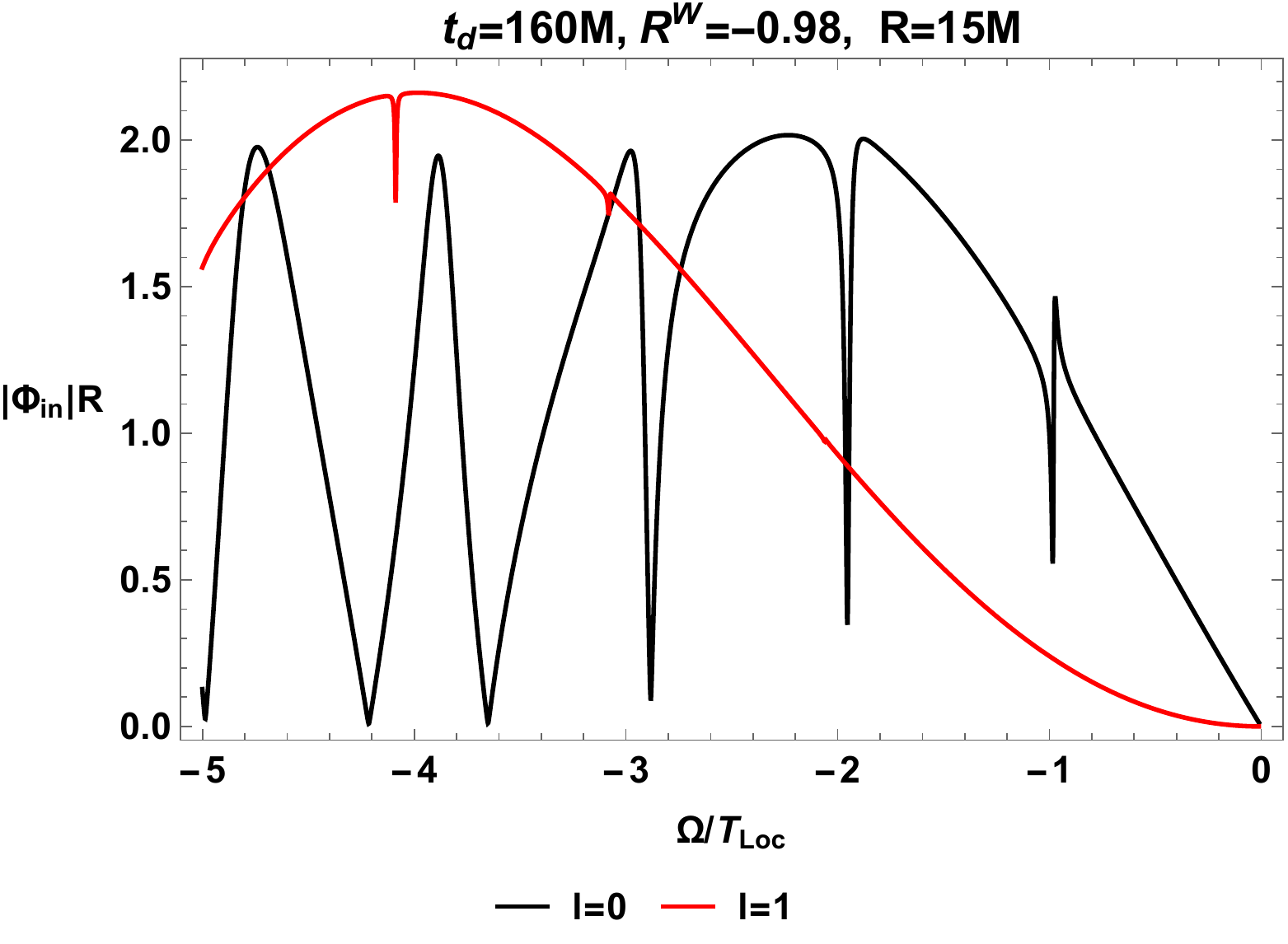}  
 \caption{(a) $\theta=\arccos(\Re(\Rin)/|\Rin|)$ and  (b)  $|\Phin| R$ ($R=15M$)  for a time delay $t_d=160M$ with $R^W=-0.98$ at $l=0$ (black) and $l=1$ (red).}
    \label{theta_Phi_td160M_RWn098}
\end{figure}

However, our numerical results indicate that the resonance heights of $\mathcal{\dot{F}_{\rm ECO}}$ for $R^W=-0.98$ are noticeably milder than those for $R^W=-1$ even at large $R$. This relates to the fact that, despite the large dips in $|\Rin|$ for  $R^W=-0.98$, it has either the same or smaller fluctuations in $\theta$  than for $R^W=-1$, as a comparison of Figure~\ref{theta_Phi_td160M_RWn098}a   with Figure~\ref{theta_Phi_td160M_RWn1}a shows. 

\section{Summary}
We have calculated the response rate of an Unruh-DeWitt detector placed away from a general exotic compact object (ECO)  having an exterior spacetime identical to that of  a Schwarzschild black hole. Due to the absence of an event horizon and the corresponding up-modes in the vacuum fields, a perfectly absorbing ECO ($R^W=0$) spacetime has a smaller detector response rate than that of the same-mass black hole, with a smaller ratio for a larger detector gap and a smaller distance. The cavity between an ECO's reflective boundary ($R^W\neq0$) and the gravitational potential peak determines the time delay $t_d$ and results in distinct resonance structures, connected to the poles of the transfer function $\Tin$. Thus, the observed spacing of the resonance structures tells us about the fundamental length scale at which the ECO deviates from a black hole.

Finally, we presented a detailed analysis on the large-distance behaviour of the resonance structures and the damping effect in the response rate, and showed how a large cavity size can increase the chances of observing resonance structure at large distances.

Our study provides a new alternative way to differentiate a static black hole from an ECO with a static UDW detector. The obvious next generalization is to 
study what happens to a spinning ECO.  Likewise, it  will be interesting to see how the results change if we consider detectors in motion, such as in geodesic orbits. We leave these investigations for future work.

\appendix
\appendixpage
\addappheadtotoc

\section{Hartle-Hawking and Unruh vacuums}
\label{sec_HH}
\begin{figure}[h]
  \centering
       \includegraphics[width=0.45\linewidth]{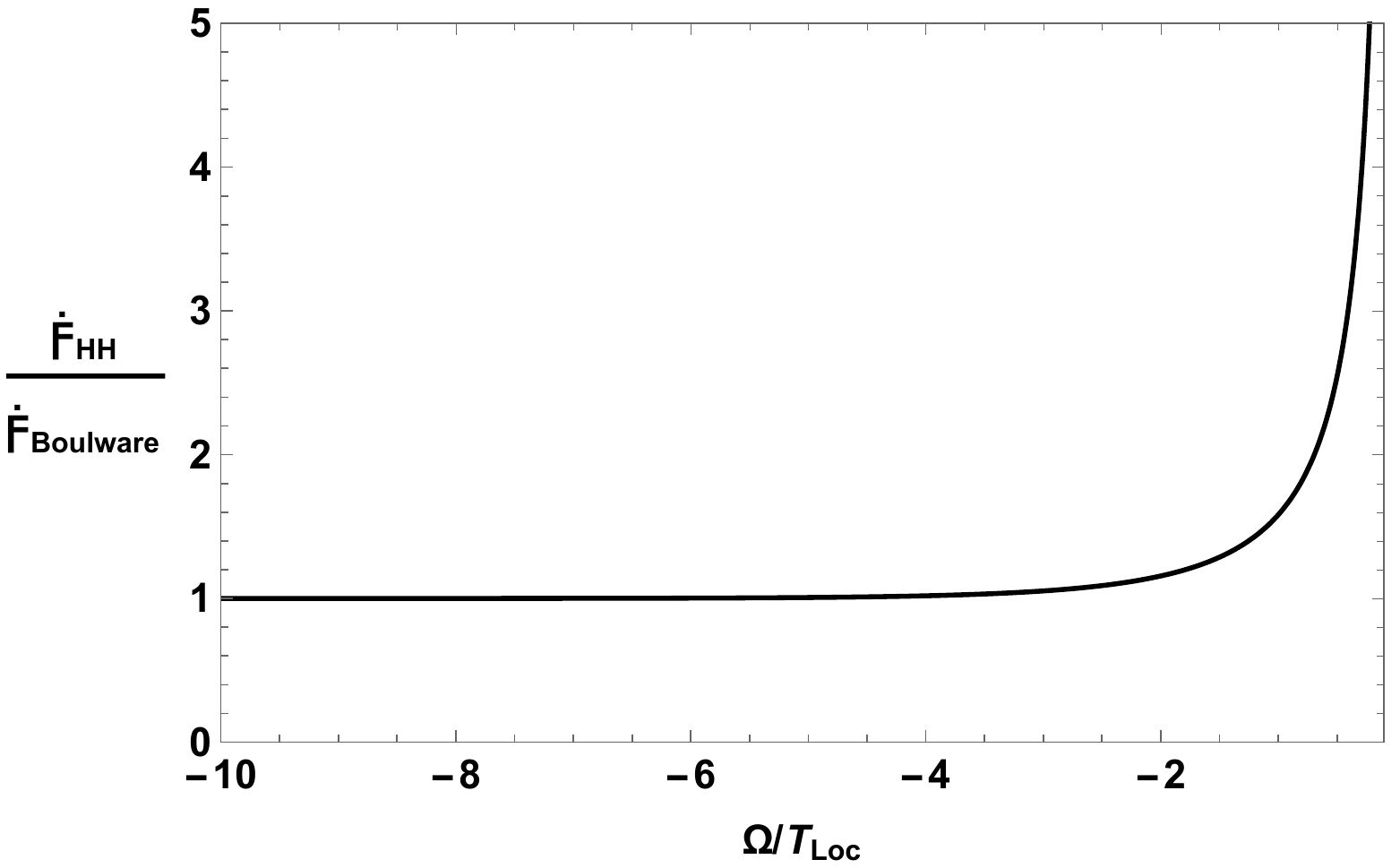}  
              \includegraphics[width=0.54\linewidth]{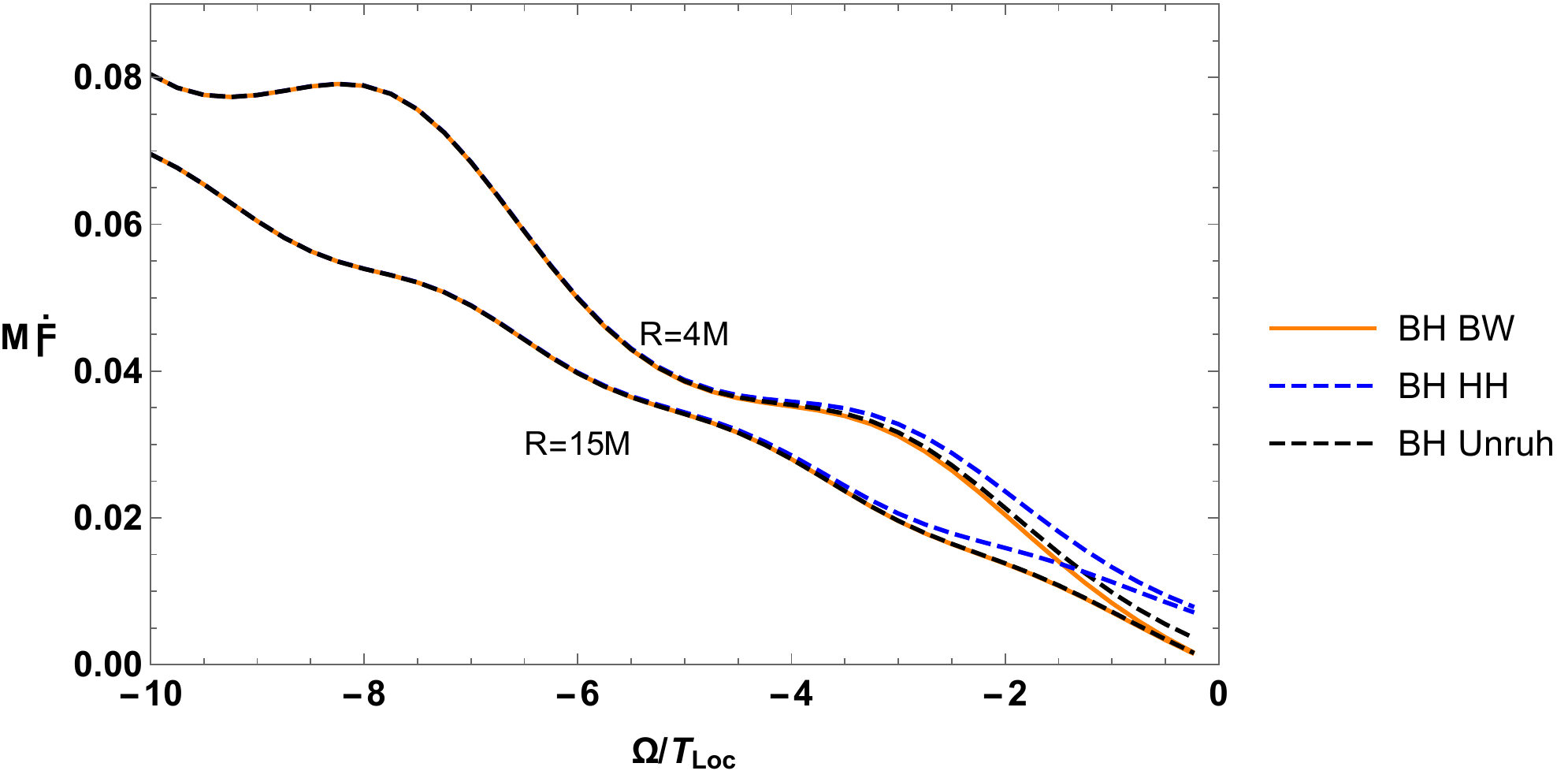}  
 \caption{(a)The ratio $ \mathcal{\dot{F}}_{HH}/\mathcal{\dot{F}}_{\rm Boulware}$ and (b) $M\mathcal{\dot{F}}_{\rm Boulware}$ (solid orange) vs   $M \mathcal{\dot{F}}_{\rm HH}$ (dashed blue) and   $M \mathcal{\dot{F}}_{\rm Unruh}$ (dashed black)  for a static detector at $R=4M$ (top) and $R=15M$ (bottom) outside a black hole.}
 \label{Fdot_HHtoB}
  \end{figure}

The Hartle-Hawking vacuum state has positive-frequency modes with respect to the horizon generators in Kruskal coordinates. It is regular across the horizon, and features outgoing and ingoing thermal fluxes from the past event horizon and the past null infinity, respectively. 
Accordingly, the response rate of  a static detector above the black hole with fields in Hartle-Hawking (HH) state are~\cite{Hodgkinson:2014iua}:
\begin{equation}
\begin{aligned}
&\mathcal{\dot{F}}_{\rm HH}\left(\Omega\right)=\frac{1}{8\pi \Omega}\frac{1}{\expo^{\Omega/T_{\text{loc}}}-1}\sum^{\infty}_{l=0}\left(2\ell+1\right)\left(|\Phi^{\text{up}}_{\tilde{\omega}\ell}(R)|^2+|\Phi^{\text{in}}_{\tilde{\omega}\ell}(R)|^2\right)\, ,
\label{Fdot_HH}
\end{aligned}
\end{equation}
Comparing to the Boulware vacuum case Eq.~(\ref{Fdot_BH_Boulware}), we have
\be
\frac{\mathcal{\dot{F}}_{\rm HH}}{\mathcal{\dot{F}}_{\rm Boulware}}=\frac{1}{\Theta(-\Omega)(1-\expo^{\Omega/T_{\text{loc}}})}, 
\label{HHtoB}
\ee
which we plot in Figure~\ref{Fdot_HHtoB}a. We see that the de-excitation rate of a UDW detector in the Hartle-Hawking vacuum becomes coincident with that of the Boulware vacuum for a large energy gap, except for an enhancement at the small energy gap region, resulting a small finite response rate at zero frequency limit ($ \mathcal{\dot{F}}_{HH}\vert_{\Omega/T_{\rm loc}\to 0}\sim 0.01$), as shown in Figure~\ref{Fdot_HHtoB}b for two benchmark examples.
  
The Unruh vacuum has positive-frequency up-modes with respect to the horizon generators in Kruskal coordinates, and positive-frequency in-modes with respect to the physical Schwarzschild  time $t$.  The response rate for a static detector with fields in Unruh vacuum state thus is~\cite{Hodgkinson:2014iua}:
\begin{equation}
\begin{aligned}
&\mathcal{\dot{F}}_{\rm Unruh}\left(\Omega\right)=\frac{1}{8\pi \Omega}\sum^{\infty}_{l=0}\left(2\ell+1\right)\left(\frac{1}{\expo^{\Omega/T_{\text{loc}}}-1} |\Phi^{\text{up}}_{\tilde{\omega}\ell}(R)|^2-\Theta(-\Omega)|\Phi^{\text{in}}_{\tilde{\omega}\ell}(R)|^2\right)\, ,
\label{Fdot_Unruh}
\end{aligned}
\end{equation}
Comparing Eq.~(\ref{Fdot_Unruh}) to Eq.~(\ref{Fdot_HH}) and  Eq.~(\ref{Fdot_BH_Boulware}), we see that the response rate for Unruh vacuum interpolates between those of Hartle-Hawking and Boulware cases. One can also see this in Figure~\ref{Fdot_HHtoB}b for two benchmark examples.


\section{Determination of $l_{\rm max}$}
\label{sec_lmax}
In section~\ref{result}, we calculated the response rate using Eq.~(\ref{Fdot_ECO_Boulware}), where the $l$ sum is taken from zero to  $l_{\rm max}$ such that the contribution with $l=l_{\rm max}+1$ is negligible.

\begin{figure}[h]
 \centering
    \includegraphics[width=0.48\linewidth]{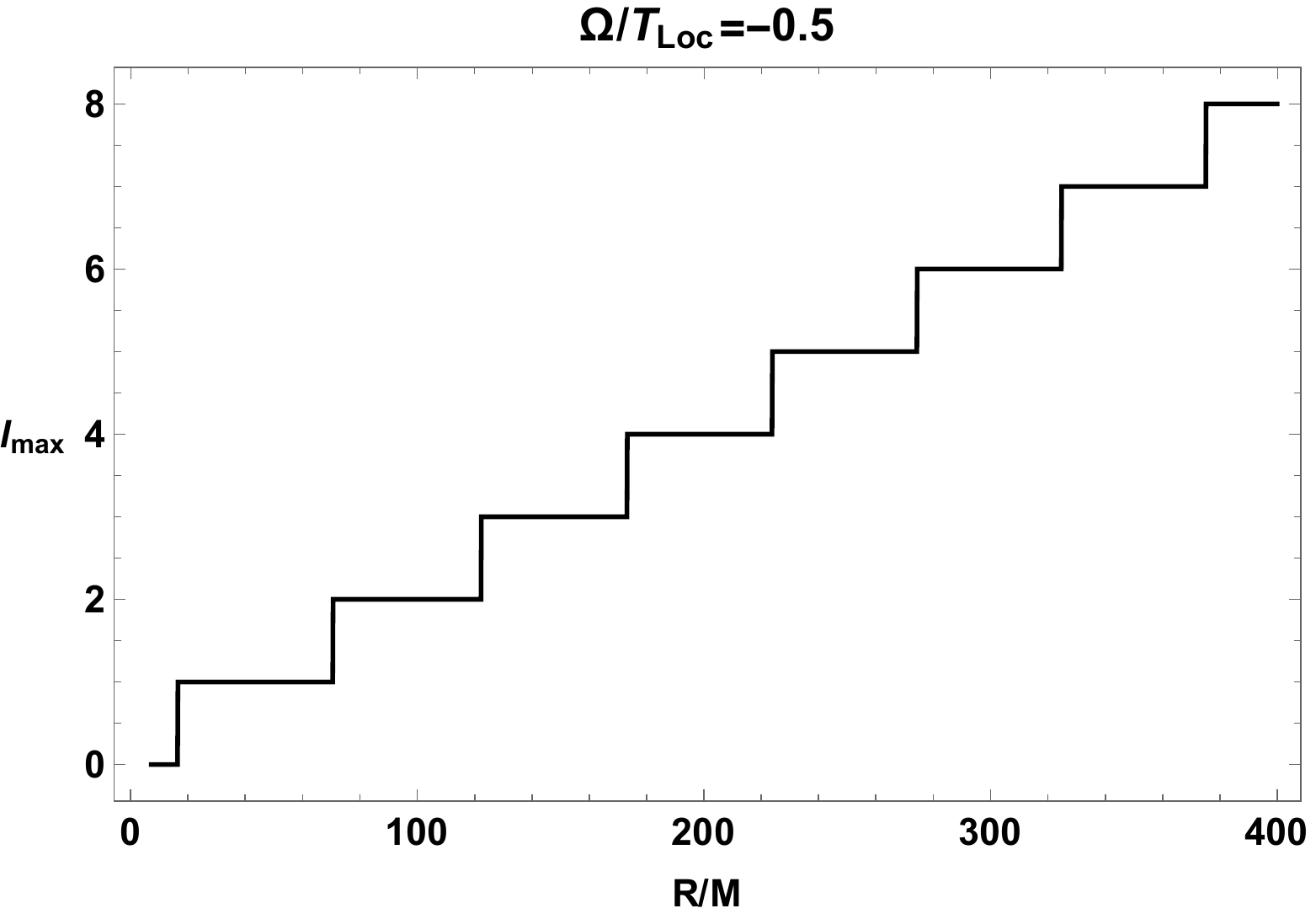}  
\includegraphics[width=0.5\linewidth]{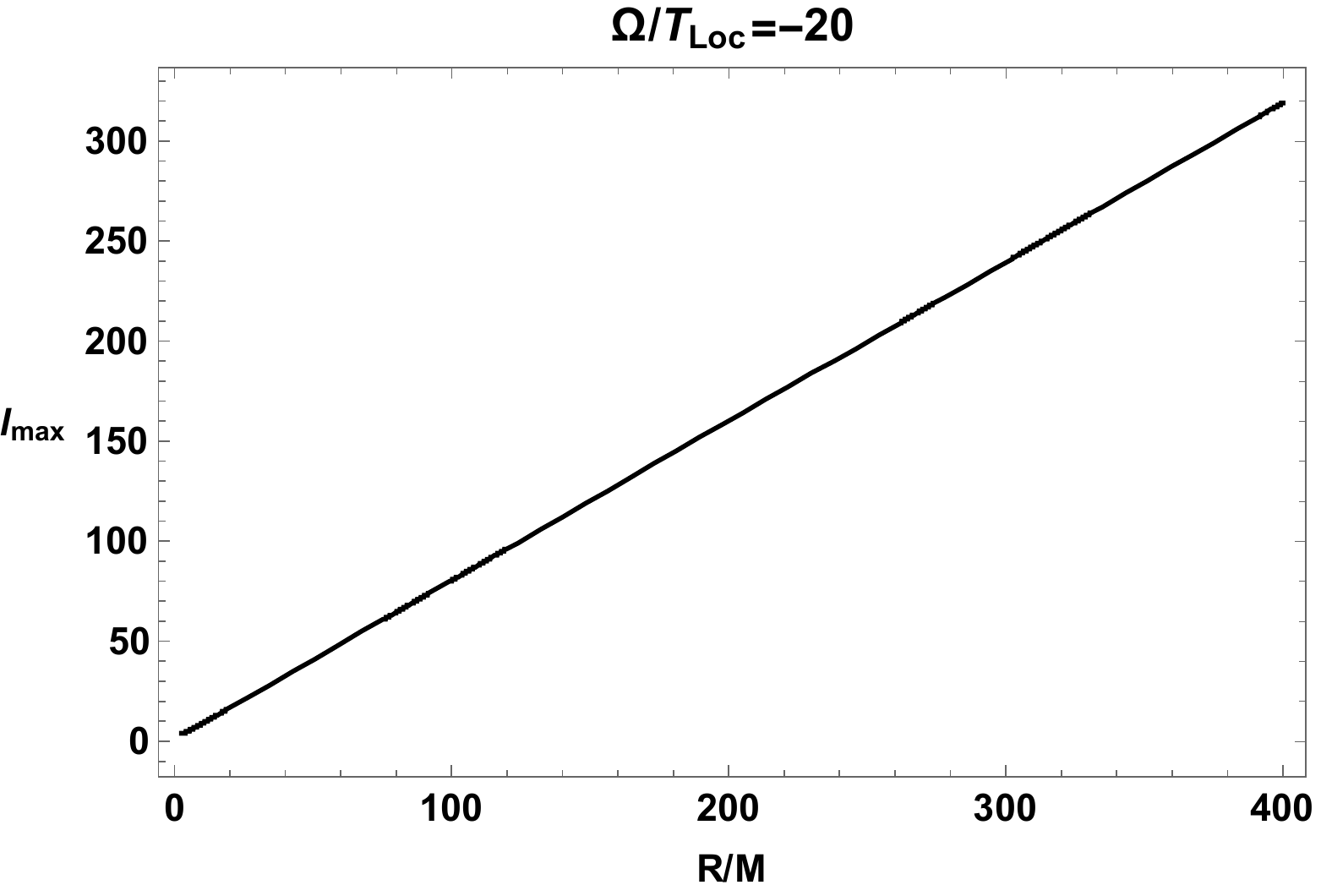}  
 \caption{$l_{\rm max}$ for a static detector at distance $R$ with (a) $\Omega/T_{\rm loc}=-0.5$ and (b) $\Omega/T_{\rm loc}=-20$.}
    \label{lmax_analy}
\end{figure}

We find that the $l_{\rm max}$ can be well approximated analytically from the relation
\be
 \omega^2-V(l_{\rm max},R)\approx0
\ee
since the wave amplitude is heavily suppressed in the region of large $(V(R)-\omega^2)$.   This gives 
\be
l_{\rm max}\approx \left\lceil \frac{1}{2} \left(-1 + \sqrt{1-\frac{8M}{R}+\frac{ 4\omega^2  R^3 }{R-2M} } \right) \right\rceil 
\ee
for given $\omega$ and $R$, where the outer bracket $\lceil\cdots\rceil$ denotes the ceiling function. Then utilizing  $\Omega/T_{\rm loc}=8\pi M \tilde{\omega}$ we obtain $l_{\rm max} (\Omega/T_{\rm loc}, R)$ accordingly, as shown in Figure~\ref{lmax_analy}. We observe the general feature that we only need to sum small $l$ values for small frequency or small $R$ region.



\section{Wronskian relations for ECO}
\label{Wronskian}
Here we derive the expressions of the transmission and reflection coefficients from the Wronskian relations.  We can construct two sets of solutions, satisfying
\begin{eqnarray}
\Phi_{\omega l}^\textrm{in}r&\to&\left\{\begin{array}{cc}
e^{-i\omega r^*_0}\,T_{\omega l}^\textrm{in}\left(e^{-i\omega(r^*-r^*_0)}+R^{\rm W}\,e^{i\omega(r^*-r^*_0)}\right), & r^*\to r^*_0 \,,  \\ 
e^{-i\omega r^*}+R_{\omega l}^\textrm{in}e^{i\omega r^*} ,& r^*\to\infty \, ,\end{array}\right.    \label{Phi_in}\\
\Phi_{\omega l}^\textrm{up} r&\to&
\left\{\begin{array}{ll}
R_{\omega l}^\textrm{up} \,e^{-i \omega r^*}+e^{i \omega r^*},&r^*\to r^*_0 \, ,  \\
T_{\omega l}^\textrm{up} \,e^{i \omega r^*},& r^*\to\infty \, , \end{array}
\right.  
\label{Phi_out}
\end{eqnarray}
Note that $\Tup=T_{\rm BH}$ and $\Rup=R_{\rm BH}$ are defined in Eq.~(\ref{Tup_norm}) for black hole with no $R^W$ dependence.

Therefore, the Wronskian satisfies
\be
W[\Phin ,\Phup]\to 
\begin{cases}
& 2i\omega\Tin (1-e^{-i 2 \omega r^*_0}R_{\rm BH} R^{\rm W})   \,, \quad\quad r^*\to r^*_0\,,\\
&2i\omega\Tup\,,\quad\quad r^*\to\infty\,.
\label{winup}
\end{cases}
\ee
Since the Wronskian must be a constant over all $r^*$, we have
\be
\Tup=\Tin (1-e^{-i 2 \omega r^*_0}R_{\rm BH} R^{\rm W}) ,
\label{tup0}
\ee
or 
\be
\Tin=\frac{T_{\rm BH}}{ 1-e^{-i 2 \omega r^*_0} {R_{\rm BH}} R^{\rm W} }.
\label{Tin_TBH}
\ee
 The resonances of $|\Tin|$ therefore result from the zeroes of $ 1-e^{-i 2 \omega r^*_0} {R_{\rm BH}} R^{\rm W}=0$. 

Since $W[\Phin,\Phup]=\Tin\Tup W[\rin,\rup],$ from Eq.~(\ref{winup}) we have
\be
\Tin=\frac{2i\omega}{ W[\rin,\rup]}.
\label{Tin}
\ee
Note that the in-mode transmission coefficient $\Tin$ is equivalent to the transfer function calculated in ECO gravitational-wave studies\footnote{Note that $\rin\leftrightarrow2i\omega \psiL$, $\rup\leftrightarrow \psiR$ for the $\psi^{\rm left,right}$ basis appearing in the gravitational wave literature~\cite{Conklin:2017lwb}. Correspondingly, $\Tin= \cal{K}$, where $\cal{K}$ is the conventional transfer function defined as ${\cal K}=1/W[\psiL,\psiR]$.  Also since $\Phin\leftrightarrow2i\omega \Tin \psiL$, $\Phup\leftrightarrow \Tup \psiR$, comparing Eq.~(\ref{Phi_in}) of this paper and Eq.~(6) and Eq. (B1) of Ref.~\cite{Conklin:2017lwb}, the coefficients map as  $ \Tin\leftrightarrow 1/(2i\omega A_{\rm in})= A_{\rm trans}/A_{\rm in}=T_{\rm BH}$, $R^{W}=R=e^{2i \omega r^*_0}\,A_{\rm ref}/A_{\rm trans} $, $\Rin \leftrightarrow A_{\rm out}/A_{\rm in}$. And $\Rup/\Tup=D_{\rm trans},\, 1/\Tup=D_{\rm ref}$, or equivalently  $\Tup=1/D_{\rm ref}=T_{\rm BH}$,  $\Rup=D_{\rm trans}/D_{\rm ref}=R_{\rm BH}$. }
\cite{Mark:2017dnq,Conklin:2017lwb}. Furthermore, it matches the $\Tin$ calculation using Eq.~(\ref{TinRinNum}) without involving any up mode.

Since
\be
W[\Phin,\Phi^{\text{up}*}_{\omega\ell}]\to 
\begin{cases}
& 2i\omega\Tin (R^{*}_{\rm BH}- e^{-2i  \omega r^*_0}R^{\rm W})   \, \quad\quad r^*\to r^*_0\,,\\
&-2i\omega T^*_{\rm BH} \Rin\,,\quad\quad\quad\quad\quad\quad\quad r^*\to\infty\,,
\label{winupc}
\end{cases}
\ee
we have
\be
\frac{W[\Phi_{\rm in},\Phi^*_{\rm up}]}{W[\Phi_{\rm in},\Phi_{\rm up}]} \Bigg|_{r^*\to \infty}=\frac{ T^{*}_{\rm BH} }{T_{\rm BH}}\frac{W[\rho_{\rm in},\rho^*_{\rm up}]}{W[\rho_{\rm in},\rho_{\rm up}]}=-\frac{ T^*_{\rm BH} }{T_{\rm BH}} \Rin,
\ee
so that 
\be
 \Rin=-\frac{W[\rho_{\rm in},\rho^*_{\rm up}]}{W[\rho_{\rm in},\rho_{\rm up}]}.
\label{RinDef}
\ee
Also, from
\be
\frac{W[\Phin,\Phi^{\text{up}*}_{\omega\ell}]^*}{W[\Phin,\Phup]} \Bigg|_{r^*\to r^*_0}=\frac{ T^{\text{in}*}_{\omega\ell} }{\Tin}\frac{W[\rin,\rho^{\text{up}\,*}_{\omega \ell}]^*}{W[\rin,\rup]}=\frac{e^{2 i  \omega r^*_0} T^{\text{in}*}_{\omega\ell} \left(-R_{\rm BH}+R^{\rm W} e^{2 i  \omega r^*_0 }\right)}{\Tin \left(-R_{\rm BH} R^{\rm W}+e^{2 i  \omega r^*_0 }\right)}
\label{Wstar}
\ee
we have
\be
\frac{W[\rin,\rho^{\text{up}\,*}_{\omega \ell}]^*}{W[\rin,\rup]}=\frac{e^{2 i  \omega r^*_0}\left(-R_{\rm BH}+R^{\rm W} e^{2 i  \omega r^*_0 }\right)}{ \left(-R_{\rm BH} R^{\rm W}+e^{2 i  \omega r^*_0 }\right)}.
\label{Wstar}
\ee
Then, defining  
\be
W^{/}=\frac{W[\rin,\rho^{\text{up}\,*}_{\omega \ell}]^*}{W[\rin,\rup]},
\label{Ws}
\ee
one has
\be
R_{\rm BH}=\frac{e^{2 i  \omega r^*_0 } R^{\rm W}-W^{/}}{1-e^{-2 i  \omega r^*_0}R^{\rm W}W^{/}}  \text{ for } R^{W}\neq \pm1
\label{rup}
\ee
and 
\be
W^{/}=\pm e^{2 i  \omega r^*_0 } \text{ for } R^{W}= \pm1
\label{Ws_pm1}
\ee from Eq.~(\ref{Wstar}).   

Referring to Eq.~(\ref{Tin}) and Eq.~(\ref{RinDef}), and the definition of $W^{/}$ (Eq.~(\ref{Ws})), we obtain the relation
\be
\Rin=\frac{\Tin}{ T^{\text{in}*}_{\omega\ell} } W^{/*}
\label{RinAsTin}
\ee
For the special case $R^W=\pm1$, inserting Eq.~(\ref{Ws_pm1}) into Eq.~(\ref{RinAsTin}), we have 
\be
\Rin=\pm\frac{\Tin}{ T^{\text{in}*}_{\omega\ell} }  e^{-2 i  \omega r^*_0 } \text{ for } R^{W}= \pm1.
\ee
For other $R^W$ value, inserting Eq.~(\ref{rup}) into Eq.~(\ref{RinAsTin}), we have
\be 
\Rin=-\frac{T_{\rm BH}(R^*_{\rm BH}-e^{-2 i  \omega r^*_0 } R^W)}{T^*_{\rm BH} (1-e^{-2i  \omega r^*_0}R_{\rm BH} R^{\rm W}) }   \text{ for } R^{W}\neq \pm1
\label{RinOfUp}
\ee
or equivalently
\be
\Rin=-\frac{T_{\rm BH}}{T^*_{\rm BH} R_{\rm BH} }\cdot\frac{|R_{\rm BH}|^2-e^{-2 i  \omega r^*_0 } R_{\rm BH} R^W}{1-e^{-2i  \omega r^*_0}R_{\rm BH} R^{\rm W}}.
\label{Rin_TBH}
\ee
Comparing Eq.~(\ref{Tin_TBH}) and Eq.~(\ref{Rin_TBH}), we see that the resonances of $\Rin$ occur at the same set of frequencies as those of $\Tin$. However, when $|R_{\rm BH}|$ goes to one (i.e., $|T_{\rm BH}|$ to 0)  at large $l$ for given $\omega$,  the second multiplicative factor's nominator cancel its denominator, making the resonance structures of $\Rin$ disappears. This helps to explain the diminishment of resonance structures for large-$l$ $\Phin$ at large-$R$. 

From Eq.~(\ref{RinOfUp}), we also see
\be
\Rin=-\frac{T_{\rm BH} R^*_{\rm BH} }{T^*_{\rm BH} } \text{ for } R^{W}= 0.
\ee
Moreover, by matching the Wronskian $W[\Phin,\Phi^{\rm in*}_{\omega \ell}]$ at $r^*_0$ and infinity we obtain the energy conservation relation $|\Tin|^2 (1-R^{W^2})=1-|\Rin|^2$. Similarly, for the Wronskian $W[\Phup,\Phi^{\rm up*}_{\omega \ell}]$,  we have $|T_{\rm BH}|^2+|R_{\rm BH}|^2=1$. 

\section*{Acknowledgements}
This work was supported in part by the  Natural Science and Engineering Research Council of Canada.




\begin{thebibliography}{}

\bibitem{Mazur:2001fv}
P.~O.~Mazur and E.~Mottola,
[arXiv:gr-qc/0109035 [gr-qc]].
\bibitem{Schunck:2003kk}
F.~E.~Schunck and E.~W.~Mielke,
Class. Quant. Grav. \textbf{20}, R301-R356 (2003)
[arXiv:0801.0307 [astro-ph]].

\bibitem{Holdom:2016nek}
B.~Holdom and J.~Ren,
Phys. Rev. D \textbf{95}, no.8, 084034 (2017)
[arXiv:1612.04889 [gr-qc]].

 
\bibitem{Baccetti:2018otf}
V.~Baccetti, R.~B.~Mann, S.~Murk and D.~R.~Terno,
Phys. Rev. D \textbf{99}, no.12, 124014 (2019)
[arXiv:1811.04495 [gr-qc]].

\bibitem{Terno:2020tsq}
D.~R.~Terno,
Phys. Rev. D \textbf{101}, no.12, 124053 (2020)
[arXiv:2003.12312 [gr-qc]].

\bibitem{Abedi:2016hgu}
J.~Abedi, H.~Dykaar and N.~Afshordi,
Phys. Rev. D \textbf{96}, no.8, 082004 (2017)
doi:10.1103/PhysRevD.96.082004
[arXiv:1612.00266 [gr-qc]].
\bibitem{Mark:2017dnq}
Z.~Mark, A.~Zimmerman, S.~M.~Du and Y.~Chen,
Phys. Rev. D \textbf{96}, no.8, 084002 (2017)
doi:10.1103/PhysRevD.96.084002
[arXiv:1706.06155 [gr-qc]].
\bibitem{Conklin:2017lwb}  
R.~S.~Conklin, B.~Holdom and J.~Ren,
Phys. Rev. D \textbf{98}, no.4, 044021 (2018)
doi:10.1103/PhysRevD.98.044021
[arXiv:1712.06517 [gr-qc]].

\bibitem{Cardoso:2019rvt}
V.~Cardoso and P.~Pani,
Living Rev. Rel. \textbf{22}, no.1, 4 (2019)
doi:10.1007/s41114-019-0020-4
[arXiv:1904.05363 [gr-qc]].

\bibitem{Holdom:2019bdv}
B.~Holdom,
Phys. Rev. D \textbf{101}, no.6, 064063 (2020)
doi:10.1103/PhysRevD.101.064063
[arXiv:1909.11801 [gr-qc]].



\bibitem{unruh} 
W.~G.~Unruh,
  Phys.\ Rev.\  D {\bf 14}, 870 (1976).
  
\bibitem{deWitt} 
B.~S. DeWitt, 
``Quantum gravity: the new synthesis'', 
in
\textit{General Relativity; an Einstein centenary survey\/} 
ed S.~W. Hawking and W.~Israel 
(Cambridge University Press, 1979) 
680. 

\bibitem{Funai:2018wqq}
N.~Funai, J.~Louko and E.~Mart\'\i{}n-Mart\'\i{}nez,
Phys. Rev. D \textbf{99}, no.6, 065014 (2019)
doi:10.1103/PhysRevD.99.065014
[arXiv:1807.08001 [quant-ph]].


\bibitem{byd} 
N.~D. Birrell 
and 
P.~C.~W. Davies, 
\textit{Quantum Fields in
Curved Space} 
(Cambridge University Press 1982).

\bibitem{Louko:2007mu}
J.~Louko and A.~Satz,
Class. Quant. Grav. \textbf{25}, 055012 (2008)
doi:10.1088/0264-9381/25/5/055012
[arXiv:0710.5671 [gr-qc]].

\bibitem{tHooft:1984kcu}
G.~'t Hooft,
Nucl. Phys. B \textbf{256}, 727-745 (1985)
doi:10.1016/0550-3213(85)90418-3


\bibitem{Hodgkinson:2014iua}
L.~Hodgkinson, J.~Louko and A.~C.~Ottewill,
Phys. Rev. D \textbf{89}, no.10, 104002 (2014)
doi:10.1103/PhysRevD.89.104002
[arXiv:1401.2667 [gr-qc]].

%


\bibitem{Holdom:2020onl}
B.~Holdom,
[arXiv:2004.11285 [gr-qc]].

\end{thebibliography}
\end{document}